\documentclass[traditabstract]{aa}
\usepackage{natbib}
\bibpunct{(}{)}{;}{a}{}{,}
\usepackage{graphicx}
\usepackage{revsymb}	
\usepackage{amsmath}	
\usepackage[usenames]{color}	
\usepackage{epstopdf}	
\newcommand{\derivp} [2] {\frac {\partial #1 } {\partial #2} }
\newcommand{\deriv} [2] {\frac {\textrm{d} #1 } {\textrm{d} #2} }

\newcommand{\eq}[1] {Eq.\,(\ref{#1})}

\usepackage{txfonts}

\usepackage[switch]{lineno} 

\begin{document}
\title{Solar $p$-mode damping rates: insight from a 3D hydrodynamical 
simulation}

\author{K. Belkacem\inst{1}, F. Kupka\inst{2,}\thanks{formerly at University of Vienna, Faculty of Mathematics, Oskar-Morgenstern-Platz 1, A-1090 Wien, Austria}, R. Samadi\inst{1}, H. Grimm-Strele\inst{3,\star}}

\institute{
LESIA, Observatoire de Paris, CNRS, PSL Research University, Universit\'e Pierre et Marie Curie, Universit\'e Denis Diderot, 92195 Meudon, France
\and
Institute for Astrophysics, Faculty of Physics, Univ. G\"ottingen, Friedrich-Hund-Platz 1, D-37077 G\"ottingen, Germany
\and 
Department of Flow and Material Simulation, Fraunhofer Institute for Industrial Mathematics ITWM, Fraunhofer-Platz 1, D-67663 Kaiserslautern, Germany
}

   \offprints{K. Belkacem}
   \mail{kevin.belkacem@obspm.fr}
   \date{\today}

  \authorrunning{Belkacem et al.}

   \abstract{
   	Space-borne missions such as CoRoT and {\it Kepler} have provided a rich harvest of high-quality photometric data for solar-like pulsators. It is now possible to measure damping rates for hundreds of main-sequence and thousands of red-giant stars with an unprecedented precision.
    However, among the seismic parameters, mode damping rates remain poorly understood and thus barely used for inferring the physical properties of stars. Previous approaches to model mode damping rates were based on mixing-length theory or a Reynolds-stress approach to model turbulent convection. While able to grasp the main physics of the problem, those approaches are of little help to provide quantitative estimates as well as a definitive answer on the relative contribution of each physical mechanism. Indeed, due to the high complexity of the turbulent flow and its interplay with the oscillations, those theories rely on many free parameters which inhibits an in-depth understanding of the problem. Our aim is thus to assess the ability of 3D hydrodynamical simulations to infer the physical mechanisms responsible for damping of solar-like oscillations. To this end, a solar high-spatial resolution and long-duration hydrodynamical 3D simulation computed with the ANTARES code allows  probing the coupling between turbulent convection and the normal modes of the simulated box. Indeed, normal modes of the simulation experience realistic driving and damping in the super-adiabatic layers of the simulation. Therefore, investigating the properties of the normal modes in the simulation provides a unique insight into the mode physics. We demonstrate that such an approach provides constraints on the solar damping rates and is able to disentangle the relative contribution related to the perturbation (by the oscillation) of the turbulent pressure, the gas pressure, the radiative flux, and the convective flux contributions. Finally, we conclude that using the normal modes of a 3D numerical simulation is possible and is potentially able to unveil the respective role of the different physical mechanisms responsible for mode damping provided the time-duration of the simulation is long enough. 
     }


   \maketitle

\section{Introduction}
\label{intro}

The \emph{Kepler} \citep{Borucki2010} and CoRoT \citep{Baglin2006a,Baglin2006b} space-borne missions provided a wealth of high-quality and long-duration photometric data which allowed us to detect thousands of   solar-like oscillating stars from the main-sequence up to the red clump \citep[e.g.][]{Chaplin2013}. A leap forward has then been achieved concerning our understanding of the internal structure of stars and their evolution \citep[e.g.][]{Mosser2012,Mosser2014,Hekker2017}. This was made possible thanks to an in-depth understanding of the physics of the oscillations \citep[e.g.][]{Belkacem06a,Belkacem06b,Belkacem2010,Dupret09,Belkacem2012,Grosjean2014,Samadi2015,Houdek2015} as well as of the evolution of the properties of the power spectra along with the stellar evolution \citep[e.g.][]{Belkacem2011,Belkacem2013}. However, since the discovery of solar five-minute oscillations, our ability to understand the physical mechanisms underlying mode damping is still challenged. Indeed, mode damping occurs in the uppermost layers of solar-like stars in which convection is highly turbulent. It also  corresponds to the location of the transition between convective and radiative energy transport. This makes the problem highly intricate because there is a concordance of several characteristic time-scales. The modal period is found to be of the same order as both the convective and thermal time-scales. Consequently, while it is notoriously difficult to model turbulent convective boundary layers \citep[see the review by][]{Kupka2017}, it is even more complex to address its coupling with the oscillations. 

Several attempts to model mode damping have nevertheless been proposed. The first to investigate this issue were \cite{Ando75} but  no stable modes were found in the whole frequency range.  \cite{Goldreich91} subsequently proposed that the shear due to Reynolds stresses, modelled by an eddy-viscosity, is of the same order of magnitude as the non-adiabatic component of the perturbation of gas pressure. \cite{Gough80} and \cite{B92a} \citep[see also][]{Houdek1999} found that the damping is dominated by the modulation of turbulent pressure, while \cite{MAD05}, \cite{MAD06c}, and \cite{Belkacem2012} also included the perturbation of the dissipation rate of kinetic energy into heat that acts to compensate the perturbation of turbulent pressure. Those formalisms were based on the mixing-length theory of convection \citep[see][for a detailed discussion]{Houdek2015}. It thus reduces the whole of the turbulent cascade to a single length-scale. Therefore, the perturbation of the mixing-length cannot account for the relation between oscillations and the turbulent cascade. \cite{Xiong00} proposed an alternative approach using a Reynolds stress formalism \citep{Canuto92} to model convection and, using a perturbation method, computed mode damping rates. However, they found a number of unstable modes, which is at odds with the observational evidence. Indeed, such an approach, while being more physically grounded than the use of the mixing-length, is highly sensitive to the adopted closure models \citep[see][for details]{Kupka2007,Kupka2007a,Kupka2007b}. 
     
Additional constraints were thus welcomed to gain more insight into the problem of mode damping. Based on the CoRoT and {\it Kepler} observations, which have provided accurate observations of solar-like oscillations of hundreds of main-sequence stars and thousands of red-giant stars, it has been shown that damping rates follow some scaling relations. Indeed, even if it is possible to reproduce the solar damping rates by tuning some parameters, it does not ensure to reproduce the damping in other stars. The observed scaling relation of mode linewidths thus provides an important additional constraint on the modeling. First, using ground-based observations, \cite{Chaplin09} proposed that mode linewidths follow a power-law dependence on effective temperature. This has been refined by \cite{Baudin2011a,Baudin2011b} using a homogeneous sample of stars observed by CoRoT and later by \cite{Appourchaux2012,Appourchaux2014} for main-sequence and sub giants as well as \cite{Vrard2018} for red giants using \emph{Kepler} observations. From a theoretical point of view, \cite{Chaplin09} predicted a power-law of $\Gamma \propto T_{\rm eff}^4$, which disagrees with CoRoT and {\it Kepler} observations. \cite{Houdek2012} attributed the failure of this theory to a missing physical mechanism and proposed mode scattering as a possible solution. However, based on the formalism developed by \cite{MAD05}, \cite{Belkacem2012} managed to reproduce the scaling relations obtained by both CoRoT and {\it Kepler}  without invoking mode scattering.   

Finally, despite some relative successes,  much work is still needed for a proper understanding of the physics behind mode linewidths. A novel approach is needed. The use of hydrodynamical 3D numerical simulation is potentially able to offer us such an opportunity. A possible way to proceed consists in constraining the free parameters of the modelling using a 3D solar simulation as proposed by \cite{Houdek2017a,Houdek2017b,Aarslev2018}. However, even though the observed damping rates are reproduced, some parameters have been tuned. Therefore, it is difficult to get insight into the physics of mode damping. A more promising approach thus consists in investigating directly the normal modes of the 3D numerical simulations. Indeed, turbulent convection generates acoustic noise  and, thanks to the boundary conditions, normal modes exist in the simulation. Even if the spatial structure of the modes is not realistic compared to the observed modes, they experience realistic driving and damping in the super-adiabatic layers of the simulation. Therefore, investigating the properties of the normal modes in the simulation provides a unique insight into the mode physics. Indeed, contrary to the observed modes, it is possible to access the perturbations associated with the oscillations in all the physical quantities and as a function of height in the simulation. For mode driving, such an approach has already been successfully used \citep[e.g.][]{Stein2001,Samadi2003}. In this article, we aim at assessing our ability to investigate mode damping using 3D numerical simulations. 
 
 This paper is organised as follows: in Sect.~\ref{simu_hydro} we present the solar 3D numerical simulation computed with the ANTARES code. The properties of the normal modes of the simulation are then described in Sect.~\ref{normal_modes}. In Sect.~\ref{work_integral}, the normal mode work integrals are computed and the contributions associated with the modulation of turbulent pressure, gas pressure, radiative and convective fluxes are investigated in Sect.~\ref{contributions_work_integral}. Finally, conclusions are provided in Sect.~\ref{conclusion}. 

\section{The solar simulation}
\label{simu_hydro}

For our subsequent analyses we use data from a numerical, hydrodynamical simulation of the solar surface. The simulation was computed as part of a more extended research project devoted to solar physics and the adequacy of numerical tools used in modelling the solar surface \citep[see][]{Grimm-Strele20015a}. One of the simulations computed for this purpose and called cosc13 was used for the present work. Its setup and some details of the simulation code are described in the following.

\subsection{The ANTARES simulation code}

The solar 3D hydrodynamical simulation cosc13 has been computed with the ANTARES code \citep{Muthsam2010} which numerically solves the Navier-Stokes equations (NSE) for a fully compressible fluid and accounts for radiative heat transfer, viscous processes, and realistic microphysics.
ANTARES uses a conservative, 5$^{\rm th}$ order weighted essentially non-oscillatory (WENO5) finite difference scheme \citep{Shu2003,Jiang1996} to discretize pressure gradients and terms representing advection in the NSE and a fourth order conservative finite difference 
scheme \citep{Happenhofer2013} to discretize dissipative terms. Time integration is done by an explicit three-stage, second order 
Runge-Kutta scheme known as SSP RK(3,2) originally proposed by \cite{Kraaijevanger1991} and analysed by \cite{Kupka2012}. \cite{Grimm-Strele2015b} demonstrated that this scheme is more efficient than traditional, total variation diminishing schemes used for this purpose (the TVD2 and TVD3 methods of \cite{Shu1988}, originally proposed by \cite{Heun1900} and \cite{Fehlberg1970}, respectively).

For cosc13 radiative transfer was treated in the non-grey approximation as described in \cite{Muthsam2010} with a multi-group technique that assigns frequencies to one of four bins, according to the optical depth at which radiation is emitted, in order to account for the full, line-blanketed spectrum in the radiative transfer. The angular integration of the radiative transfer equation required to compute the radiative flux was performed using a three-point Gauss-Radau integration along polar coordinates per hemisphere and a four-point equispaced integration along the azimuthal direction. 
Using a short-characteristics method \citep[cf.][]{Muthsam2010} the scheme hence features 18 ray directions including two vertical rays into each hemisphere. The latter was not the case for the scheme used by \cite{Muthsam2010}. As described in the latter the radiative source and sink term, $\vec \nabla \cdot \vec F_{rad}$, is treated in a stationary approximation, since changes in the simulation occur on time-scales that are orders of magnitude longer than the light crossing time of a unity optical depth \citep[see, e.g., ][Chapts. 6.5 and 7.2]{Mihalas1984}. 
The transition between the surface layers, for which the radiative transfer equation is solved, and the lower lying layers, where the diffusion approximation holds, is obtained from smooth interpolation between the two solutions for grid cells always located in the optically thick region: the average temperature there is typically around 12500 K and thus the mean optical Rosseland depth is in a range from about 1000 to several 1000, so even for extreme events optical thickness is ensured \citep[cf. also Fig.~15 of][]{Stein98}. 

\subsection{Setup of the simulation: microphysics and simulation grid}

The sum of gas and radiative pressure, the thermodynamical derivatives, and related thermodynamical quantities are obtained from interpolating 
in the LLNL equation of state tables \citep{Rogers1996}. Realistic (radiative) conductivities are obtained from interpolation in the opacity 
data of \cite{IglesiasRogers1996} for interior layers. Non-grey opacities are obtained from the tables of \cite{KuruczCD2,KuruczCD13} with 
the binning procedure described in \cite{Muthsam2010} where also further details on the construction of opacities and equation of state 
from these tables are given \citep[see][too]{Grimm-Strele20015a}. The old standard solar composition of \cite{GN93} was assumed.

The numerical simulation has been conducted for a box with a Cartesian grid containing a volume of 3.88 Mm as measured vertically and 6 Mm 
as measured horizontally. The spatial location of this ``box within the Sun'' was chosen to contain the solar surface such that layers with an 
optical Rosseland depth larger than $10^{-4}$ always remained inside the simulation box. Open boundaries as in \cite{Grimm-Strele20015a} 
(type BC3b, from their Table 1) have been used in vertical directions as well as periodical boundary  conditions for both horizontal 
directions. A uniform resolution has been chosen with a grid spacing of 11.1 km from top to bottom for cells which are 35.3 km wide. 
The computations for this model have been done on the VSC-2 (Vienna Scientific Cluster) on 144 CPU cores in parallel. MPI parallelization 
was used and due to extra grid cells required to implement the boundary conditions the total grid actually contained 359 grid points in the vertical 
and 179 points along the two horizontal directions. From this grid 350 vertical layers can be extracted which consist of 170 by 170 cells 
that are used for computing horizontal averages during post-processing of the simulation.

\subsection{Initial conditions, model relaxation, and statistical evaluation}


The simulation was relaxed from its initial condition for about $5884\,$s or $18.58$ sound crossing times.  The latter are measured from the top of the simulation box to its bottom for the initial model. In the present case this has been the helioseismologically calibrated standard solar model S \citep{modelS}. Since that model is actually a bit too shallow at the top we extended the simulation domain upwards  by about 200 km. In \cite{Grimm-Strele20015a} the procedures to do so have been explained. It has also been shown there that the specific 1D starting model is not crucial, as simulations started from different solar structure models which agree in effective temperature and input entropy in the quasi-adiabatic layers near the bottom all yield the same mean stratification after relaxing the simulation with respect to kinetic energy, i.e., after about one solar hour. 

\begin{figure}[t]
\begin{center}
\includegraphics[width=9.8cm]{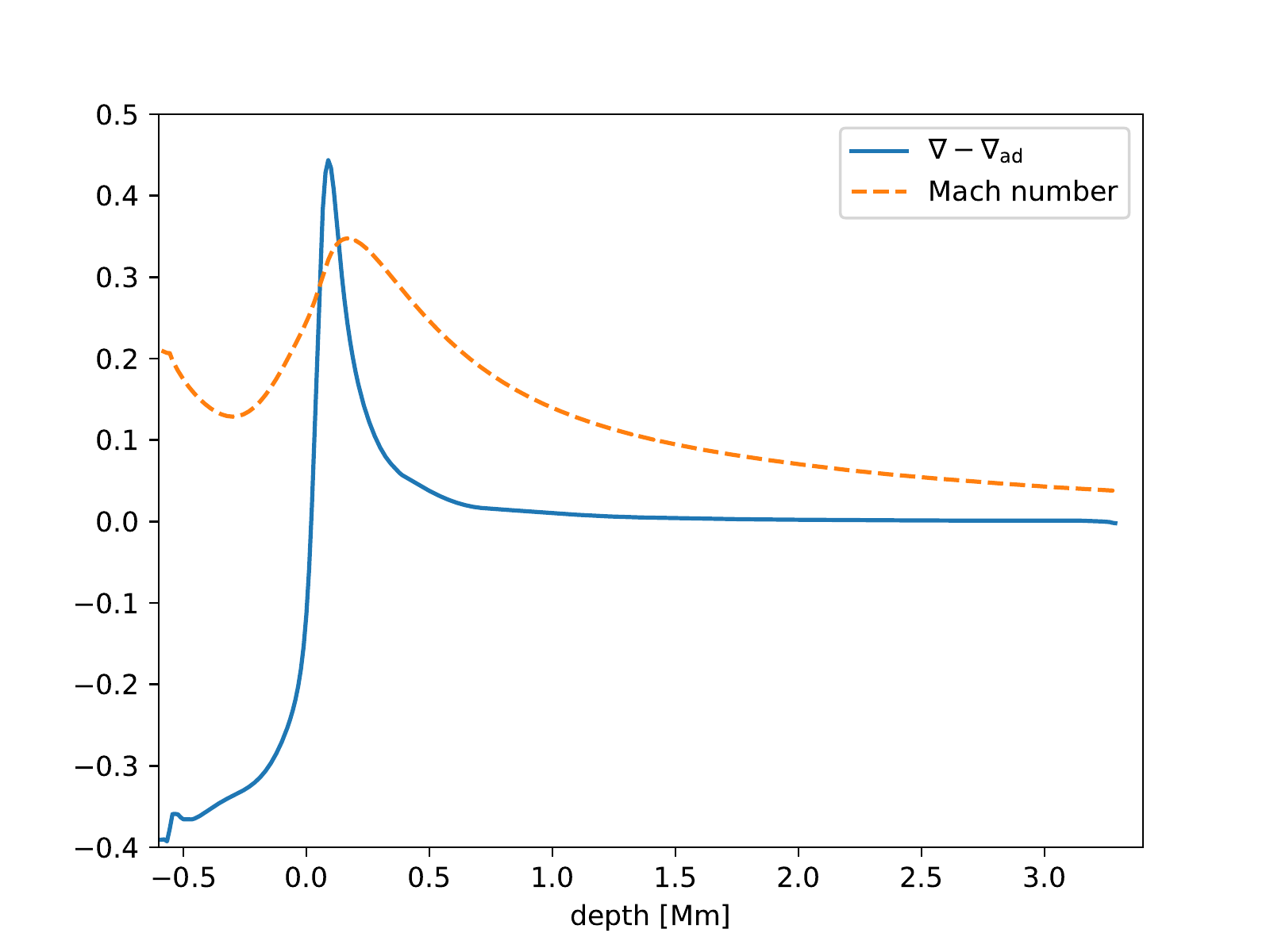}
\caption{The super-adiabatic gradient ($\nabla - \nabla_{\rm ad}$) and the Mach number (defined as the ratio between the temporally and horizontally averaged convective vertical velocity and  sound speed) versus the simulation depth. The zero-point depth is chosen where the temporally and horizontally averaged temperature equals the effective temperature. \label{numbers}}
\end{center}
\end{figure}

The relaxed simulation was then evaluated every 1/20 of a sound crossing time, i.e.\ at an interval of 15.84 sec. This way 2527 samples have 
been produced covering a time of about 40012~sec or slightly more than 11 hours. The samples are not strictly equidistant in time, 
 since the  samples are picked from the dynamically varying time-stepping of the simulation  (variations introduced this way are in any case less than 
about 0.2 sec and most of the time below 0.1 sec). Statistical quantities have been computed for this output in a post processing step to calculate 
both horizontally averaged variables for each output step, i.e.\ for each of the 2527 samples, as well as ensembles averages from the horizontally 
averages quantities over the entire integration time of more than 126.3 sound crossing times. The data generated this way was used in the analysis 
presented in the following and the difference between the temperature gradient and the adiabatic gradient, as well as the Mach number, are displayed in Fig.~\ref{numbers} as a function of depth in the simulation. 
From averaging the vertical component of the radiative flux at the surface over the entire simulation time 
an effective temperature can be determined for the numerical simulation. Over the entire simulation time  $T_{\rm eff}$ is found 
to be on average $5750~{\rm K} \pm 18~{\rm K}$, with a negligibly small drift of $-1.1~{\rm K}$ per hour over that time and rare extrema not exceeding $52~{\rm K}$. The drift has been computed from a least square fit of a linear function to $T_{\rm eff}$ as determined at 0.1 Mm below the top of the simulation, where $F_{\rm tot}=F_{\rm rad}$ within less than 0.1\%. Visual inspection and a comparison with a straight mean, which is more than $1.6~{\rm K}$ lower than the mean of the least square fit line, confirm that the drift becomes smaller with time and it is in any case well within the range of fluctuations of $T_{\rm eff}$ determined for our simulation. We hence conclude the simulation to show an agreement with solar effective temperature. A value for the effective temperature 
of the Sun which is commonly used in recent literature is $5779$~K. It can be derived from the results discussed in \cite{JCD2009} and is also provided in Table 18.2 of \cite{CoxGiuli68}. However, values differing from this result by a few K depending on the specific measurements for radius and luminosity used are also common \citep[see, e.g.,][]{Lebreton2008}. This is certainly sufficiently accurate to investigate solar p modes. The surface gravity naturally remains fixed
at its initial value of $\log(g)=4.4377$ as does the chemical composition, $(X,Y,Z)=(0.7373, 0.2427, 0.0200)$,
with its \cite{GN93} metallicity distribution.

\section{Normal radial modes of the simulation}
\label{normal_modes}

 The first step is to identify the normal modes of the simulation and thus to determine their characteristics throughout the simulation. To this end, we consider radial modes only so that, as described in Sect.~\ref{simu_hydro}, we consider horizontal averages of the simulation for the physical quantities. 
 
\subsection{Mode profiles and fitting procedure}
\label{profile}

\begin{figure}[t]
\begin{center}
\includegraphics[width=9.2cm]{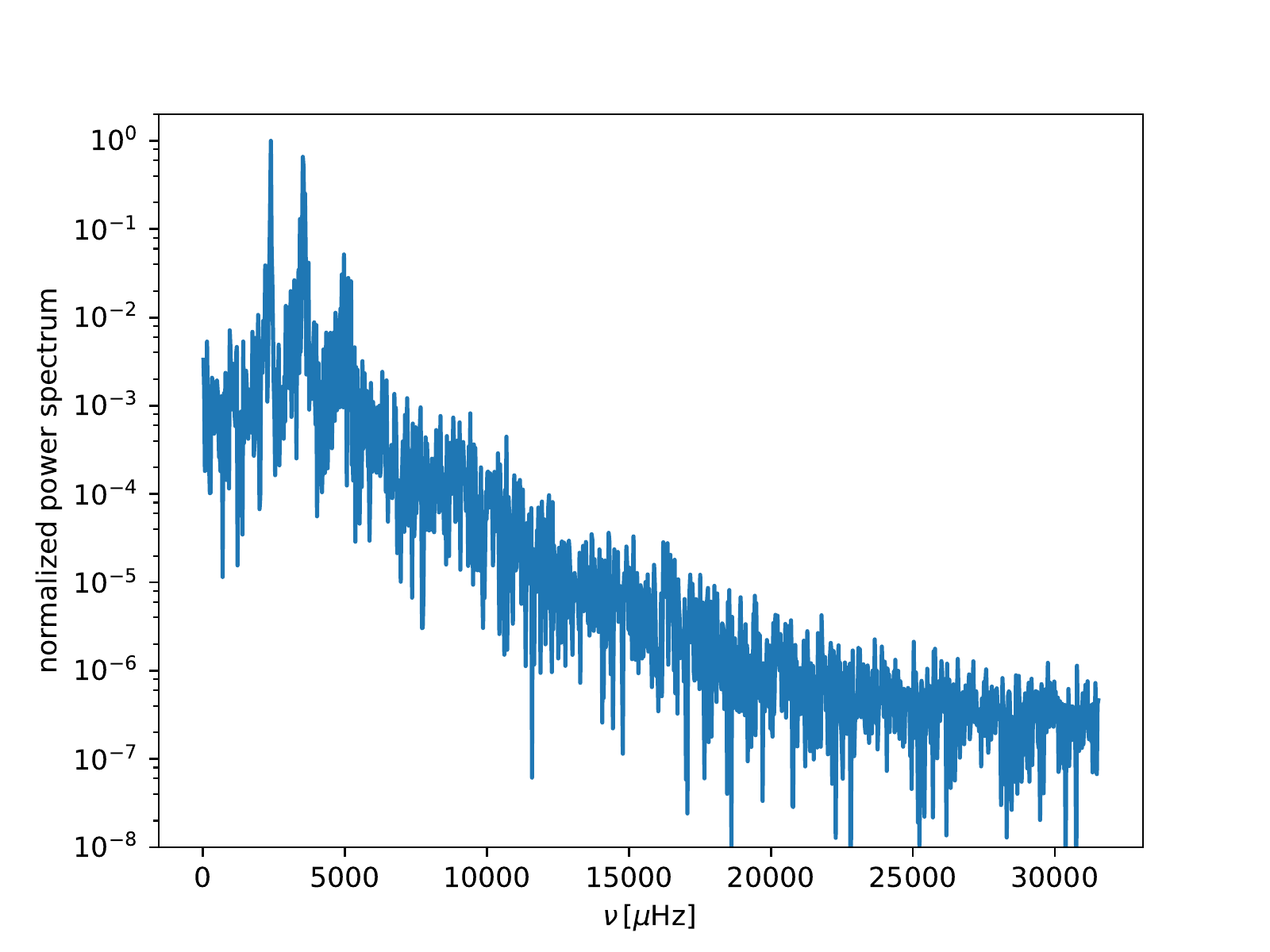}
\caption{Power spectrum of the vertical velocity at the photosphere, normalized by its maximum value, as a function of the frequency. \label{power_vz}}
\end{center}
\end{figure}

We Fourier transform the time-series described in Sect.~\ref{simu_hydro} using a Fast Fourier Transform 
algorithm. As shown by Fig.~\ref{power_vz}, one can clearly distinguish three normal modes with a Lorentzian profile that is characteristic 
for solar-like oscillations.
Indeed, in the time series, the vertical velocity of a radial solar-like mode can be written as
\begin{align}
\label{mode_time_domain}
{\rm v}_{\rm osc} (\omega_0) = \omega_0 \, A \, \xi_r(r) \, \cos \left(\omega_0 t - \phi \right) \, e^{-\eta t} \, , 
\end{align}
where $A$ is the $A$ is  the amplitude at $t=0$. The observed signal is a sum of many such terms, each with their own amplitude and zero-point of time, and also phase. 
 $\xi_r$ is the radial displacement eigen-function, $\omega_0=2\pi \nu_0$ is the pulsational eigen-frequency, $t$ is the time, $\eta$ is the damping rate, and $\phi$ the phase. 

In the power spectrum, for $\nu \approx \nu_0$, the Fourier transform of \eq{mode_time_domain} can be approximated by a Lorentzian function such as \citep[e.g.][]{Baudin2005,Appourchaux2014}
\begin{align}
\label{eq:oscillator:fourier2}
\left\vert \widehat{{\rm v}}_{\rm osc} \right\vert^2 = \frac{H}{1+x^2} \,, \quad {\rm with}\quad x=2(\nu-\nu_0)/\Gamma \, , 
\end{align}
and $H$ stands for the mode height, $\Gamma$ is the mode linewidth that is related to the mode damping rate through $\Gamma = \eta / \pi$. Mode height is subsequently related to the mode amplitude and mode linewidth by \citep[e.g.][]{Samadi2011}
\begin{align}
{\rm v}_{\rm osc}^2  & = \pi \, H \, \Gamma \, .
\label{v_s_2} 
\end{align}
Therefore, a normal mode in the Fourier spectrum can be characterized by several quantities. First, through the global quantities 
(that do not vary with depth), \emph{i.e.} frequency and mode linewidth. Second, through the mode height and phase which 
depend on the location in the simulation. 

\begin{table}
   \centering
    \caption{Global characteristics of the three normal modes as displayed in Fig.~\ref{power_vz}, where $\nu_0$ is the central 
    frequency, $\Delta \nu_0$ is the error on the frequency, $\Gamma$ is the mode linewidth, and $\Delta \Gamma^+, \Delta \Gamma^-$ are 
    the upper and lower errors on the mode linewidth, respectively. Note that the first mode (mode 1) is not resolved so that 
    the only constraint we get is that its linewidth is lower than the frequency resolution, \emph{i.e.} $\Gamma \le 25\, \mu$Hz. }
  \begin{tabular}{lccccc} 
      \hline
      Mode    & $\nu_0$ [$\mu$Hz]& $\Delta \nu_0$ [$\mu$Hz]& $\Gamma$ [$\mu$Hz]& $\Delta \Gamma^{+}$ [$\mu$Hz]& $\Delta \Gamma^{-}$ [$\mu$Hz]\\
      \hline
       1      & 2397.95 &  0.42 & - & -  & -\\ 
       2 & 3540.41  & 0.98  & 50.00 & 2.17 & 2.08\\
       3 & 4955.14  & 4.39  & 319.28 & 14.08 & 13.48\\      
      \hline
   \end{tabular}
   \label{tab:mods1}
\end{table}

We then fitted the power spectrum of the vertical component of the velocity by means of the maximum-likelihood estimator \citep[see e.g.][]{Toutain1994,Appourchaux1998}. 
Each mode is fitted separately using a constant background. The global parameters, \emph{i.e.} mode frequencies and linewidths, are obtained by performing a simultaneous fitting in several layers. For the fundamental mode of the simulation (hereafter mode 1), we consider all layers except near the upper and bottom boundaries. 

The results for the frequencies and linewidths are summarized in Table~\ref{tab:mods1}.\footnote{Internal errors are computed from the Hessian matrix as described in \cite{numrecipies}.} We note that the linewidth of mode 1 is not provided since it is lower than the actual resolution (which is about $25 \, \mu$Hz). We also emphasize that  although the values of the global parameters in Table~\ref{tab:mods1} are precise, due to the relatively short duration of the simulation, it is difficult to obtain accurate results. Indeed, as shown for instance by \cite{Appourchaux2014}, the determination of mode linewidth is subject to many biases. For example, an  overestimation of  the mode height leads to an underestimation of the linewidth. 
Therefore, the values provided in Table~\ref{tab:mods1} should be considered with care and are only intended to provide order of magnitudes. 

\begin{figure}[t]
\begin{center}
\includegraphics[width=9.8cm]{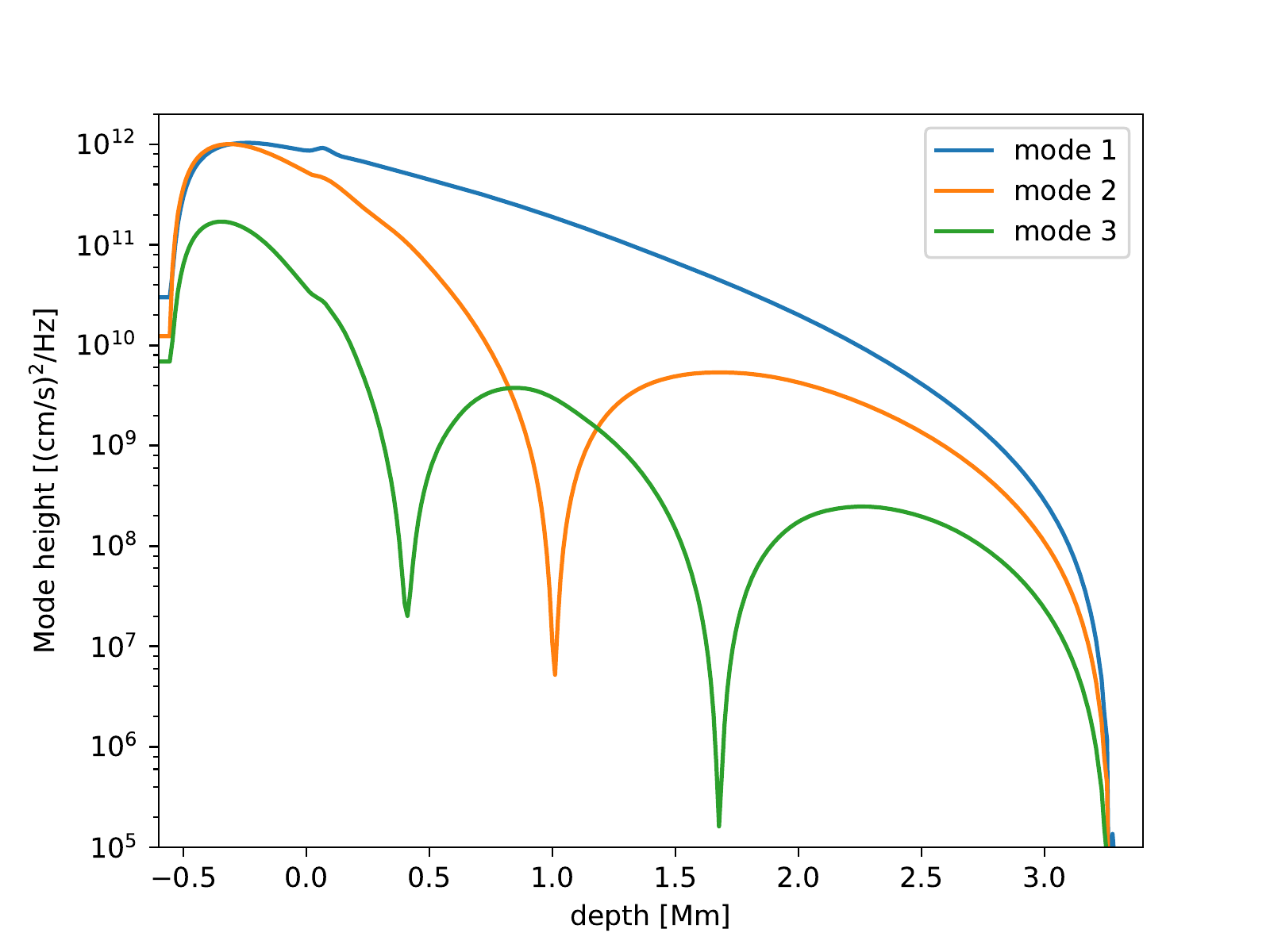}
\includegraphics[width=9.8cm]{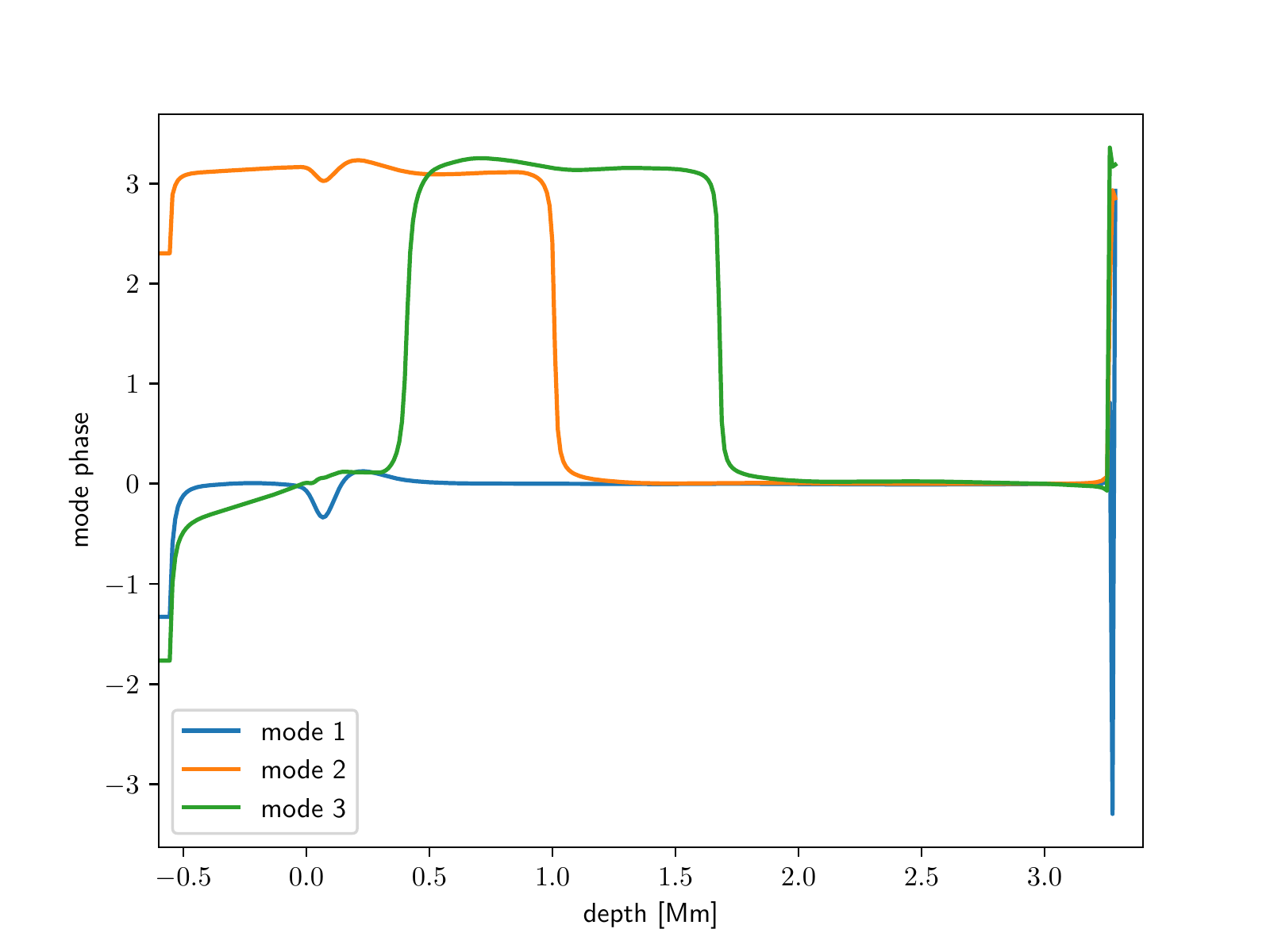}
\caption{\emph{Top panel}: Mode  velocity power density as a function of the depth for the three modes identified in Table.~\ref{tab:mods1}.  As for Fig.~\ref{numbers}, the zero-point depth is chosen where the temporally and horizontally averaged temperature equals the effective temperature.  \emph{Bottom panel}: The same as for the top panel except that the mode phases, computed directly in the Fourier space, are displayed.  The zero-point phases are chosen at a depth equals to $3$\, Mm. \label{height_vz}}
\end{center}
\end{figure}

To obtain the mode height and phase we consider the frequency bin with the largest power, near the eigen-frequency. 
We note that  inferring those quantities  from fits at each layer within the simulation is possible, but it provides much more noisy results.  
Figure~\ref{height_vz} displays the mode  velocity power densities (top panel) and mode phases (bottom panel) as a function of depth. The first mode corresponds to the fundamental mode with no node, while the second and third mode exhibit one and two nodes, respectively. We note that both for the mode velocity power densities and phases, there is a rapid variation near the peak of the super-adiabatic gradient which is typical for non-adiabatic effects. Indeed, this feature is the result of a rapid variation of the entropy perturbations \citep[e.g][]{Belkacem2011,Samadi2012}. 

\subsection{Comparison with adiabatic oscillations of a solar 1D model}
\label{identification_modes}

We can even go a step further and identify the normal modes of the simulation with the observed modes. Indeed, as already shown by \cite{Stein2001}, it is possible to compare and identify the 3D modes with modes computed with a standard 1D model. To this end, one has to identify modes that exhibit a node at the bottom of the simulation and compare the mode velocity profiles. To do so, the velocity of the normal modes of the simulation is computed in the Fourier domain using \eq{v_s_2} for the amplitude. Note that for the first mode, we used the spectral resolution instead of the linewidth since it is not resolved. 

For consistency, we thus compute a 1D solar model  that matches both the solar gravity and effective temperature but also the mean temperature at the bottom of the 3D numerical simulation. This model has been obtained using the CESTAM evolutionary code \citep{Marques2013} assuming standard physics: Convection was included according to \cite{CGM1996}, with a mixing-length parameter $\alpha=0.67$, and turbulent pressure is ignored. 
Microscopic diffusion was included. The OPAL equation of state is assumed. The chemical mixture of the heavy elements is similar to that  of \citet{GN93}'s mixture. Subsequently, we constructed a 1D model following \citet{Trampedach97} as detailed in \citet{Samadi08,Samadi2010b} in such a way that their outer layers are replaced by the averaged 3D simulations.  Finally, 1D adiabatic oscillations are computed using the ADIPLS code \citep{JCD11} and the "gas $\Gamma_1$'' hypothesis to account for the turbulent pressure \citep[see][]{Rosenthal1999,Sonoi2015,Sonoi2017}.

\begin{figure}[t]
\begin{center}
\includegraphics[width=9.8cm]{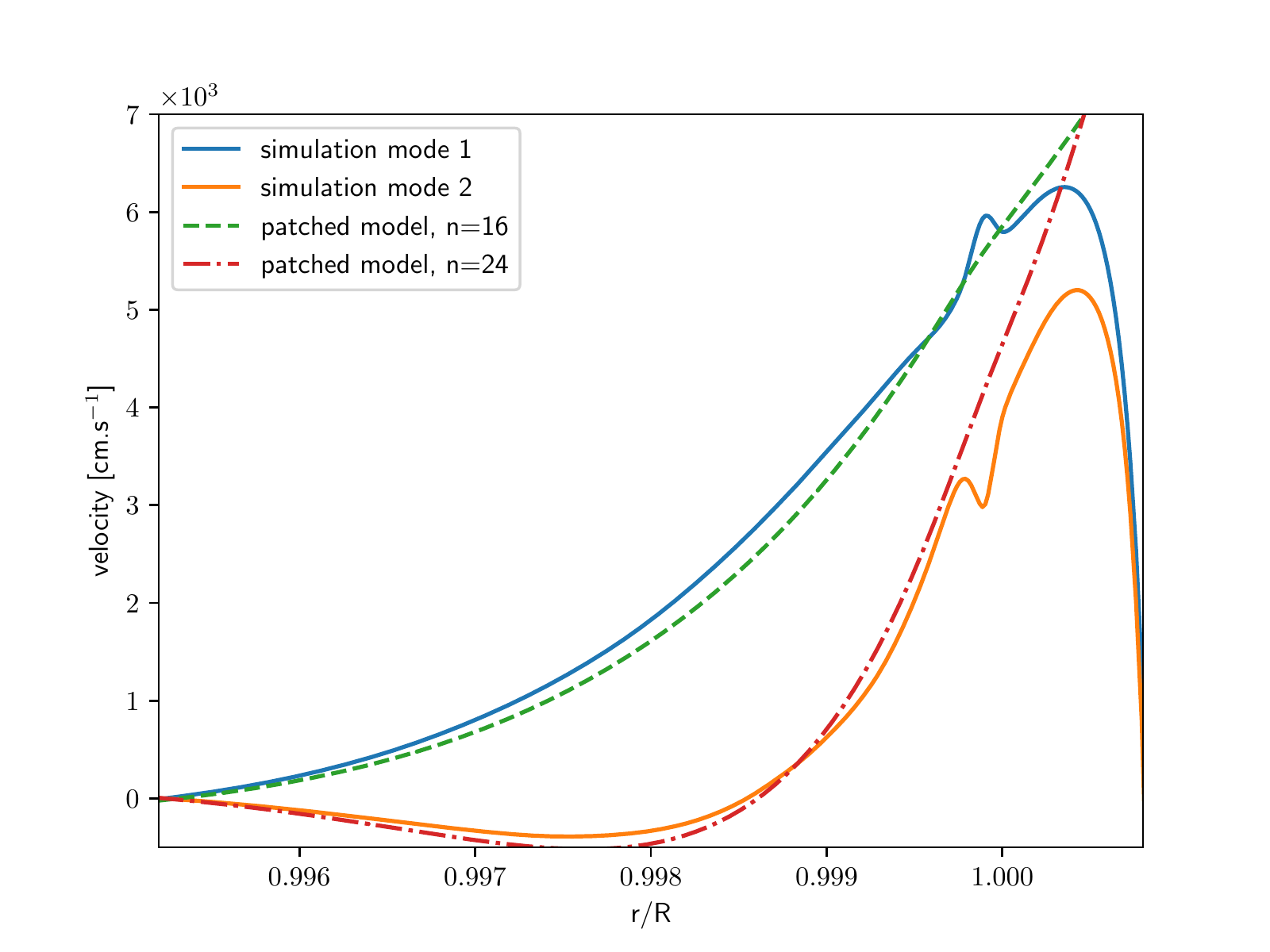}
\includegraphics[width=9.8cm]{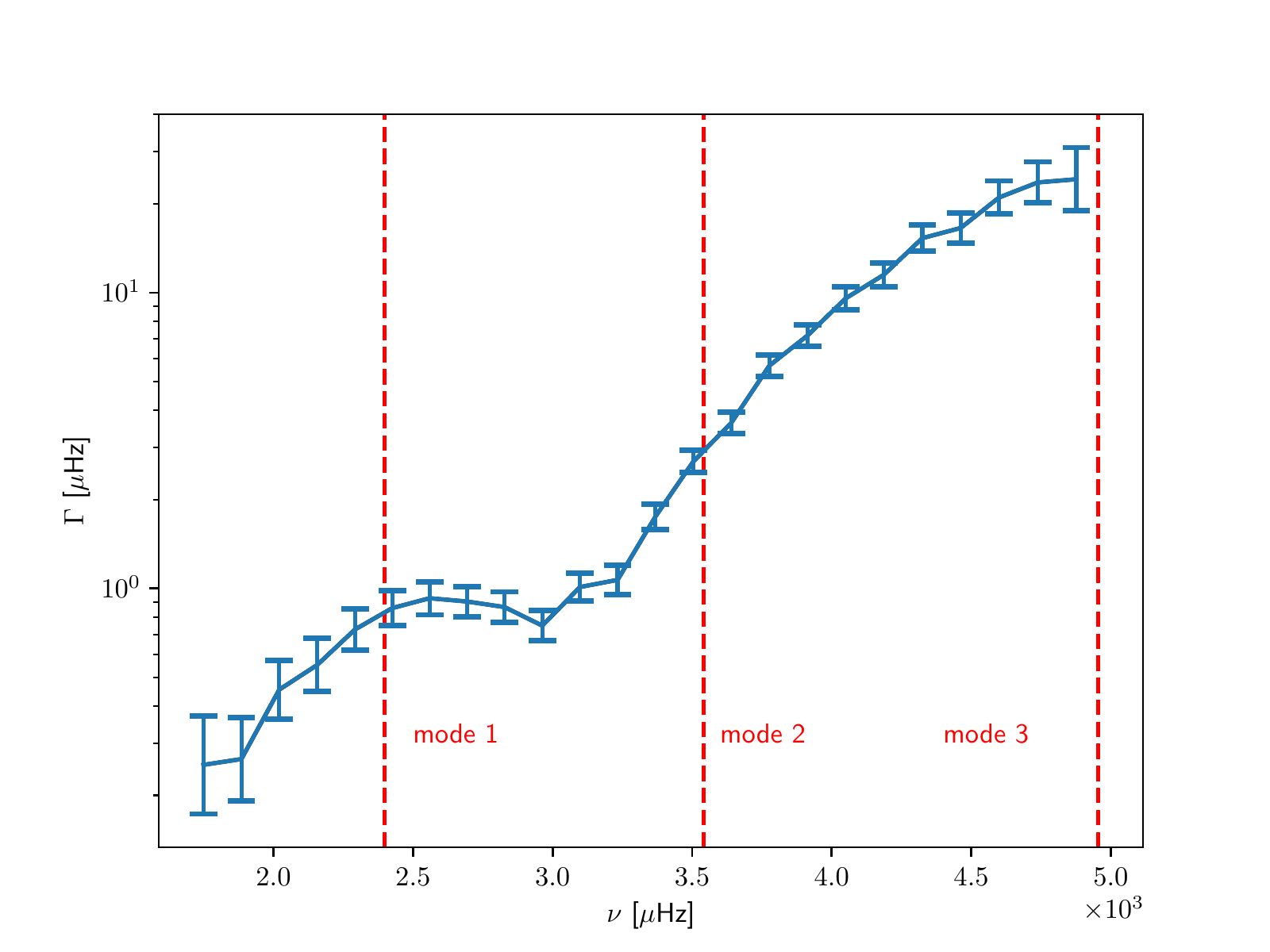}
\caption{\emph{Top panel:} Mode velocity as a function of the  radius normalized by the total radius of the Sun obtained using the 1D solar model ( i.e., where the temperature equals the effective temperature). For the normal modes of the 3D simulation, the velocity profiles have been obtained as described in Sect.~\ref{profile} and using \eq{v_s_2}. For the normal modes computed using the 1D solar model, the velocities are obtained as described in Sect.~\ref{velocity_depth} and they are normalized so that their kinetic energy equals the kinetic energy of the corresponding mode in the simulation. \emph{Bottom panel:} Mode damping rates observed by the GOLF instrument as a function of frequency. The observations are taken from \cite{Baudin2005}. The over-plotted red vertical dashed lines correspond to the frequencies of the modes in the 3D simulation (see Table.~\ref{tab:mods1}) and are used to emphasize the correspondence with the observed solar modes. \label{identification}}
\end{center}
\end{figure}

Figure~\ref{identification} (top panel) shows the comparison of normalized velocities and the resulting mode identification for modes 1 and 2. The eigen-velocities from the 1D computation have been normalized so that their kinetic energy equals the kinetic energy of the corresponding mode in the 3D simulation. It turns out that the three modes of the 3D simulation correspond to the observed modes with respective radial orders $n=16$, $n=24$, and $n=33$. There is a very good  match between the velocity profiles of both modes 1 and 2 as a function of radius compared to the adiabatic 1D computation. The main differences occur in the atmospheric layers since the upper boundary of the 1D model and the 3D simulation are different as well as near the peak of the super-adiabatic gradient (\emph{i.e.} at the bottom of the photosphere). In the latter region, the sharp variations exhibited by the 3D modes are the result of purely non-adiabatic effects so that adiabatic computations are unable to reproduce those features. As it will be shown in the following sections, these patterns are important for investigating the physics of mode damping.  In the uppermost region, the difference between the 1D and 3D models can be attributed to the boundary condition of the 3D simulation, which forces a node for the normal modes.

In Fig.~\ref{identification} (bottom panel), we illustrate the mode identification by showing the solar damping rates as a function of frequency. The modes that correspond to the normal modes of the simulation are over-plotted in red and it appears that modes 1 and 2 of the 3D simulation  bracket  the frequency of the maximum height ($\nu_{\rm max} = 3100\, \mu$Hz) while mode 3 corresponds to a mode near the cut-off frequency. 
 Since the relative importance of the various physical contributions to the damping rates is expected to vary with frequency \citep[e.g.][]{B92a,Belkacem2012}, this will allow us to probe different physical regimes with respect to the mode damping. 
We also note that the frequencies of the modes in the simulation and the frequencies of the observed modes are comparable but not exactly the same. This is a consequence of the location of the bottom of the simulation and, to a lesser extent, to non-adiabatic effects \citep[e.g.][]{Sonoi2017}. 

\subsection{Scaling for mode amplitudes}
\label{velocity_depth}

Going beyond the shape of the eigenfunctions, one can expect that the mode physics in the simulation is realistic enough to gain some insight into the physical mechanisms responsible for mode damping. Indeed, mode damping occurs in the upper-most layers of the stars and more precisely when the thermal time-scale becomes equal or higher than the modal period \cite[see for instance][]{Belkacem2011,Belkacem2012}. This occurs in the super-adiabatic layers and in the atmosphere, but in the quasi-adiabatic layers, mode damping and driving are negligible. Consequently, the extent of the simulation appears to be sufficient to investigate the physical mechanisms responsible for mode damping. 
 However, at first sight, the normal modes of the 3D numerical simulation cannot be directly compared to the observed solar oscillations or to the computed solar oscillations from a 1D model. The fundamental difference is the size of the resonant cavity, which modifies the mode masses (see Eq.~\ref{def_mode_mass}) as well as the large separation.  Indeed, while for the Sun the large separation is $\Delta \nu_{\mathrm{sun}} \simeq 135 \, \mu$Hz, we found $\Delta \nu_{\mathrm{3D}} \simeq 1930 \, \mu$Hz for the 3D simulation by computing the first-order asymptotic expression of the large separation \citep[e.g.][]{Tassoul1980}.

Using order of magnitude estimates, it is nevertheless possible to obtain a relation between the mode amplitude in the 3D simulation and in the Sun. To do so, let us first define the energy of a mode as \citep[e.g.][]{Samadi2011,Belkacem2013,Samadi2015}
\begin{align}
\label{mode_energy}
E_{\mathrm{osc}} = \mathcal{M} \, {\rm v}_{\mathrm{osc,s}}^2  \, ,
\end{align}
where ${\rm v}_{\mathrm{osc,s}}$ is the mode velocity observed at the surface of the Sun and $\mathcal{M}$ the mode mass defined as  
\begin{align}
\label{def_mode_mass}
\mathcal{M} = \frac{1}{\vert {\rm v}_{\mathrm{osc,s}}\vert^2} \int \, \vert {\rm v}_{\mathrm{osc}} \vert^2 \, {\rm d}m \, , 
\end{align}
where ${\rm v}_{\mathrm{osc}}$ is the eigen-velocity. 

To go further, we note that the eigen-velocity amplitude becomes important near the surface layers so that its integration over the vertical coordinate is similar within the 3D box and in the 1D model. 
 More precisely, using the 1D model, the relative contribution of the mode mass in the upper most layers corresponding to the 3D simulation domain is about $11\%$, $8\%$, and $7\%$ for modes 1, 2, and 3 respectively. Consequently, in the 3D and 1D models, they mainly differ due to the horizontal integral, so that the mode masses per unit surface area are approximately similar in the Sun and in the 3D simulation. Therefore
\begin{align}
\label{equality_normalized_mode_masses}
\frac{\mathcal{M}_{\mathrm{3D}}}{S} \simeq \frac{\mathcal{M}_{\mathrm{Sun}}}{4 \pi R^2} \, ,
\end{align}
with
\begin{align}
\mathcal{M}_{\mathrm{3D}} = \frac{S}{\vert {\rm v}_{\mathrm{osc,s,3D}} \vert^2} \int {\rm d}z \, \rho_0 \, \vert {\rm v}_{\mathrm{osc,3D}} \vert^2 \, , 
\end{align}
and 
\begin{align}
\mathcal{M}_{\mathrm{Sun}} = \frac{4\pi R^2}{\vert {\rm v}_{\mathrm{osc,s,Sun}} \vert^2} \int {\rm d}r \,\left(\frac{r}{R}\right)^2 \rho_0 \, \vert {\rm v}_{\mathrm{osc,Sun}} \vert^2 \, , 
\end{align}
where $R$ is the solar radius, $S$ is the horizontal surface of the simulation, $\rho_0$ is the horizontally and time-averaged density, $z$ the depth of the 3D simulation, 
and the subscripts $\mathrm{3D}$ and $\mathrm{Sun}$ stand for the modes of the 3D simulation and the solar modes, respectively. By considering the second mode of the 3D simulation (which is resolved and has a sufficient signal to noise ratio) and the corresponding mode in our solar 1D model (see Sect.~\ref{identification_modes}), we found that \eq{equality_normalized_mode_masses} is verified to a few percent. 

Now using \eq{mode_energy} together with \eq{equality_normalized_mode_masses}, one obtains 
\begin{align}
\label{relation_energies}
E_{\mathrm{osc,3D}} \simeq E_{\mathrm{osc,Sun}} \, \left(\frac{S}{4 \pi R^2}\right) \, \left(\frac{v_{\mathrm{osc,s,3D}}}{v_{\mathrm{osc,s,Sun}}}\right)^2 \, .
\end{align}
We also note that the mode energy can be written as \citep[see][for details]{Samadi2015}
\begin{align}
E_{\mathrm{osc}} = \frac{\mathcal{P}}{2 \eta} \, , 
\end{align}
where $\mathcal{P}$ is the excitation rate. Consequently, mode energy is independent of the mode mass (or mode inertia) since both the excitation and damping rates can be written to be inversely proportional to the mode mass. Now, if the 3D simulation is realistic enough, one can hence expect that the energies of a mode in the 3D simulation and its equivalent in the Sun are similar (\emph{i.e.} $E_{\mathrm{osc,3D}} \simeq E_{\mathrm{osc,Sun}}$). We checked this relation by computing directly the mode energies in the 3D simulation and in the solar model using \eq{mode_energy}. For the 3D simulation, the mode velocity at the surface is computed directly using the Fourier transform of the velocity and for the solar surface velocity we consider $v_{\mathrm{osc,s,Sun}}$ from \cite{Baudin2005}. It gives, using \eq{v_s_2} and \eq{mode_energy}, $E_{\mathrm{osc,3D}} \simeq 2.5 \times 10^{19}$J and  $E_{\mathrm{osc,Sun}} \simeq 1.9 \times 10^{20}$J (for $\ell=1$), which is in  a reasonable agreement given the uncertainties of our fit\footnote{We also note that the observation made with the GOLF instrument is obtained at an altitude much higher than the photosphere. Therefore, the mode energy of the observed mode at the photosphere is expected to be lower \citep[see][at the frequency of the maximum power]{Kjeldsen2008}.} (see Sect.~\ref{profile}). 

Finally, using the aforementioned argument and using \eq{relation_energies}, we get 
\begin{align}
\label{ratio_vitesses_final}
\frac{{\rm v}_{\mathrm{osc,s,3D}}}{{\rm v}_{\mathrm{osc,s,Sun}}}  \simeq \left(\frac{\mathcal{M}_{\mathrm{Sun}}}{\mathcal{M}_{\mathrm{3D}}}\right)^{1/2} 
\simeq \left(\frac{4 \pi R^2}{S}\right)^{1/2} \simeq 400 \, .
\end{align}
Solar mode amplitudes are typically between $0.1$ m.s$^{-1}$ and $0.3$ m.s$^{-1}$ near the frequency of the maximum amplitude and using \eq{ratio_vitesses_final} one recovers the order of magnitude of the amplitudes of the normal modes in the 3D simulation, which are about $100$ m.s$^{-1}$.  
Obviously, this estimate yields a very rough order of magnitude and differences in mode physics between the 3D model and the observations are still to be expected, for instance, due to the boundary conditions of the simulation. Nevertheless, this estimate favors the idea that mode damping and mode driving occurring within the 3D simulation is worth being considered for constraining the underlying physical processes. 

\section{Computation of the work integral}
\label{work_integral}

Mode damping can be directly inferred by fitting the modes in the power spectrum of the vertical velocity. However, if one wants to go further and decipher the contributions to this damping, it is necessary to compute the work integral. Indeed, such an approach permits us to explicitly split the contributions but also to infer information on the location of both driving and damping regions in the simulations. 

\subsection{Averaged equations}

The first step consists in averaging the primitive equations. To do so, as we are considering a compressible flow, we will consider both Reynolds and mass weighted averages  \citep[e.g.][]{Canuto1997,Nordlund2001}. In the following, we will  approximate the Reynolds average by the straight horizontal average in the 3D simulation. Therefore, any quantity $X$, can be decomposed such that
\begin{align}
\overline{X} = X - X^\prime \, , \quad \mathrm{with} \quad \overline{X^\prime} = 0 \, .
\end{align}
This average will be applied to the density,  pressure, radiative and convective fluxes.  

The second average is named as the mass weighted or Favre average \citep{Favre1969}, so that for a quantity $X$, it is defined as
\begin{align}
\widetilde{X} = \frac{\overline{ \rho X}}{\overline{ \rho }} \, .
\end{align}
 The quantity $X$ can thus be decomposed such that  
\begin{align}
X = \widetilde{X} + X^{\prime\prime} \,  .
\end{align}
It immediately follows
\begin{align}
\overline{\rho X^{\prime\prime}} = 0 \, ,  \quad \mathrm{and} \quad 
\overline{ X^{\prime\prime}} \neq 0 \, .
\end{align}
A more detailed description of the properties of the Favre average is provided in Appendix.~\ref{defs_averages}. Notice that in our case, the mass average will be applied to the velocity, and entropy fields \citep[see][for details]{Canuto1997,Nordlund2001}.  Because we consider a compressible flow, this choice permits us to simplify the equations. Indeed, many correlation terms involving density fluctuations are incorporated into the Favre mean quantities and consequently are no longer present in the governing equations by using the Favre average for the above-mentioned quantities. 

Subsequently, we average the mass conservation equation (see Appendix.~\ref{mean_mass_conv} for a detailed derivation) to obtain
\begin{align}
	\label{mass_wave}
	&\deriv{\overline{\rho}}{t} + \overline{\rho}\, \derivp{\widetilde{u}_z}{z} = 0 \, , 
\end{align}
where $\rho$ is the density, $u_z$ is the vertical component of the velocity. We also introduced the pseudo-Lagrangian derivative such that
\begin{align}
\deriv{}{t} = \derivp{}{t} + \widetilde{u}_z \derivp{}{z} \, .
\end{align}
Applying the same procedure,  and omitting the terms that cancel from hydrostatic equilibrium, we get from the momentum conservation equation (see Appendix.~\ref{mean_mass_conv} for details)
\begin{align}
	\label{moment_wave}
	&\overline{\rho} \deriv{\widetilde{u}_z}{t} = -\derivp{\delta P_g}{z} -\derivp{\delta P_t}{z} - \delta \rho \, g \, , 
\end{align}
where $P_t$ is the turbulent pressure, $P_g$ is the gas pressure, $g$ is the gravitational field that is considered constant as in the 3D simulation. In addition, for any quantity $X$, we have defined $\delta X \equiv \overline{X}-\langle\overline{X}\rangle_t$ with $\langle\overline{X}\rangle_t$ the time average of $\overline{X}$. $\delta X$ is therefore the pseudo-Lagrangian perturbation of $X$. Finally, we introduced the notation $X_0 \equiv \langle\overline{X}\rangle_t$ .

\subsection{Integral expression of the damping rates}

To get insight into the physics of mode damping, it is necessary to determine the phase lag between the Lagrangian perturbations of pressure and density for a given mode 
within the simulation and the integral of this phase lag provides the total damping rate for that mode \cite[e.g.][]{Samadi2015}. This is what we call the integral approach. Therefore, 
we will work in the time Fourier space. To start, we thus consider the time Fourier transformation of \eq{moment_wave} and we multiply it by $\rho_0 \widehat{u_z}^\ast$ to obtain 
\begin{align}
\label{momentum_wave_fourier2}
i \omega \rho_0 \left\vert \widehat{u_z}\right\vert^2 = 
- \widehat{u_z}^\ast \widehat{\frac{\rho_0}{\overline{\rho}} \derivp{\delta P}{z}} 
- \widehat{u_z}^\ast \widehat{\frac{\delta \rho}{\overline{\rho}}} \rho_0 g \, , 
\end{align}
where $\delta P = \delta P_g + \delta P_t$ is the total fluctuation of pressure, the symbol $(\; \hat{} \;)$ stands for the  temporal Fourier transform, the symbol $(\; ^\ast \;)$ stands for the complex conjugate,  and $\omega$ is the cyclic frequency.\footnote{Notice that $\omega$ (the cyclic frequency in the time Fourier domain) is not to be confused with $\omega_0$ (the modal frequency).} Note that the tilde has been omitted for ease of notation.
This equation can be further simplified if one keeps only the dominant order in the LHS of \eq{moment_wave}. This is equivalent to assuming $\overline{\rho} \simeq \rho_0$ in the RHS of \eq{momentum_wave_fourier2} such that 
\begin{align}
\label{momentum_wave_fourier3}
i \omega \left\vert \widehat{u_z}\right\vert^2 = 
- \frac{\widehat{u_z}^\ast}{\rho_0} \derivp{\widehat{\delta P}}{z} 
- \widehat{u_z}^\ast \widehat{\frac{\delta \rho}{\rho_0}} g \, .
\end{align}

To go further, we integrate over the mean mass column density $\tau_0 \equiv \left<\tau\right>_t$ (defined as ${\rm d}\tau_0 = \rho_0 {\rm d}z $), perform integration by part and multiply by $-i\omega$ to finally obtain
\begin{align}
\label{momentum_tmp2}
\omega^2 = 
\frac{i\omega}{E} \int {\rm d} \tau_0 \left\{ - \frac{\widehat{\delta P}}{\rho_0} \derivp{\widehat{u_z}^\ast}{z} 
+ \frac{1}{\rho_0} \derivp{\left(\widehat{\delta P} \widehat{u_z}^\ast\right)}{z} 
+ \widehat{u_z}^\ast \widehat{\frac{\delta \rho}{\rho_0}} g \right\} \, , 
\end{align}
where  
\begin{align}
E = \int {\rm d}\tau_0 \left\vert \widehat{u_z}\right\vert^2 \, .
\end{align}
Note that at the modal frequencies, $E$ can be identified with the mode inertia. 

It is now useful to take advantage of the Fourier transform of the mass conservation equation that reads 
\begin{align}
\label{fourier_continuity}
i \omega \frac{\widehat{\delta \rho}}{\rho_0}= - \derivp{\widehat{u_z}}{z} \, , 
\end{align}
where again $\overline{\rho} \simeq \rho_0$, so that \eq{momentum_tmp2} leads to 
\begin{align}
\label{work_integral_form1}
&\eta = \frac{\mathcal{I}m(\omega^2)}{2\omega_R} = 
\frac{1}{2 \omega_R E} \int {\rm d} \tau_0  \, \left(\mathcal{T}_1 + \mathcal{T}_2 + \mathcal{T}_3 \right)
\end{align}
where 
\begin{align}
\mathcal{T}_1 &\equiv \vert \omega \vert^2 \mathcal{I}m \left(\frac{\widehat{\delta P}}{\rho_0} \frac{\widehat{\delta \rho}^\ast}{\rho_0} \right) \, , \\
\mathcal{T}_2 &\equiv \mathcal{R}e\left(\frac{\omega}{\rho_0} \derivp{\left(\widehat{\delta P} \widehat{u_z}^\ast\right)}{z} \right) \, ,\\
\mathcal{T}_3 &\equiv - \mathcal{I}m \left(\widehat{u_z}^\ast \derivp{\widehat{u_z}}{z}  g\right) \, .
\end{align}
We assumed that near the modal frequency (\emph{i.e.} $\omega \simeq \omega_0$) the coherent contributions (associated with the oscillation) of the pressure and density fluctuations dominate over the random contributions (associated with the turbulence). 
The eigen-frequency is complex $\sigma_0 = \sigma_{\rm R} + i\eta$, with $\eta$ the damping rate, and in \eq{work_integral_form1} we assumed $|\sigma_0|^2 = \sigma_{\rm R}^2 + \eta^2
\simeq \sigma_{\rm R}^2$ since we are in the situation for which $\sigma_{\rm R} \gg \eta$. 

At this step, several comments are necessary. The second term of \eq{work_integral_form1}, i.e. $\mathcal{T}_2$, vanishes because $\widehat{u_z}^\ast$ is null at the bottom of the simulation box and both $\widehat{u_z}^\ast$ and $\widehat{\delta P}$ tend to zero at the upper boundary. We checked numerically that this contribution is negligible  as well as the contribution of the third term ($\mathcal{T}_3$). Consequently, \eq{work_integral_form1} reduces to
\begin{align}
\label{work_integral_form2}
\eta = \frac{\omega_R}{2 E} \int {\rm d} \tau_0  \, \mathcal{I}m \left(\frac{\widehat{\delta P}}{\rho_0} \frac{\widehat{\delta \rho}^\ast}{\rho_0} \right) \, .
\end{align}
Thus, the damping rate is determined by the phase lag between mode compressibility and  pressure fluctuations. This is a classical result in 1D non-adiabatic calculations \citep[e.g.][]{Ledoux1958,Unno89}. 

An alternative approach, as proposed by \cite{Nordlund2001}, is to split the pressure fluctuations in an adiabatic and non-adiabatic component, i.e.
\begin{align}
\label{split_pressure_nad}
\delta P = \delta P^{\rm ad} + \delta P^{\rm nad} \, , 
\end{align}
where 
\begin{align}
\label{GGM}
\delta P^{\rm ad} &= c_{s,0}^2 \, \delta \rho \, ,
\end{align}
with $c_{s,0}^2$ the horizontal and time averaged squared sound speed \citep[see Appendix in][]{Samadi07}. The adiabatic pressure fluctuations are dominant over the non-adiabatic ones but do not contribute to the damping. This can be easily seen by inserting \eq{split_pressure_nad} into \eq{work_integral_form2}. Consequently, splitting explicitly the adiabatic and non-adiabatic contribution can help to improve the accuracy of the computation. However, we checked that using either the total or the non-adiabatic pressure fluctuations makes almost no difference when computing \eq{work_integral_form2}. 

\subsection{Computation of the cumulative damping}

We then computed \eq{work_integral_form2}. To do so, the pseudo-Lagrangian quantities ($\delta X$) are computed by interpolating the physical variables ($X$) onto the time-averaged mean column density ($\tau_0$) before subtracting their time-averaged values. Then, the amplitudes and phases of each quantity ($\delta X$) are taken from the Fourier transforms. Eventually, the damping rates are obtained at the frequency of the modes. More precisely;
\begin{itemize}
	\item For the first mode, since it is not resolved, we consider the bin at the maximum height of the mode. The result is $\eta_1 = 5.13 \, \mu$Hz, which is consistent with our upper limit (\emph{i.e.} the resolution of $25 \, \mu$Hz) because of the relatively  short  duration of the simulation. 
	\item For the second mode, we adopt the same approach by selecting the bin with the highest amplitude in the Fourier spectrum because this bin is the least affected by the noise. 
	We find $\eta_2 = 59.75\,\mu$Hz, which is quite close to the value found from fitting the mode-peak (see Table~\ref{tab:mods1}).  
	\item For the third mode, the situation is more complex because of its large width. In this case, and due to the low signal to noise ratio, $\eta$ is highly varying from bin to bin preventing us from making conclusions.
\end{itemize}
Finally, given the inherent limitations of the simulation, one can conclude that for the first two modes the damping rates computed from 
Eq.~(\ref{work_integral_form2}) and the measured damping rates are roughly consistent (see Table~\ref{tab:mods}). However, for the third one, it is not possible to draw conclusions. Indeed, one can easily expect large uncertainties on the fitted mode linewidths because of the relatively short duration of the simulation compared to the fifteen years of continuous observation of the Sun by the GOLF instrument. The same is true for the damping rate computed using the integral expression (Eq.~\ref{work_integral_form2}) because near the bottom of the simulated box the imaginary part of mode perturbations become smaller than stochastic convective fluctuations so that the estimate of the phase lag between the non-adiabatic pressure and the density is highly affected by the noise. 

\begin{table}
	\centering
	\caption{Global characteristics of the three first normal modes, where $\nu_0$ is the central 
		frequency and $\Gamma$ is the mode linewidth (see text for details).}
	\begin{tabular}{lccc} 
		\hline
		Mode    & $\nu_0$ [$\mu$Hz]& $\Gamma$ (fit) [$\mu$Hz]& $\Gamma$ (work integral) [$\mu$Hz]\\
		\hline
		 1      & 2397.95 &  $\leq$ 25 & 5.13 \\
		 2 & 3540.41  & 50.00  & 59.75 \\
		 3 & 4955.14  & 319.28  & -  \\      
		\hline
	\end{tabular}
	\label{tab:mods}
\end{table}

\begin{figure}[t]
	\begin{center}
		\includegraphics[width=9.8cm]{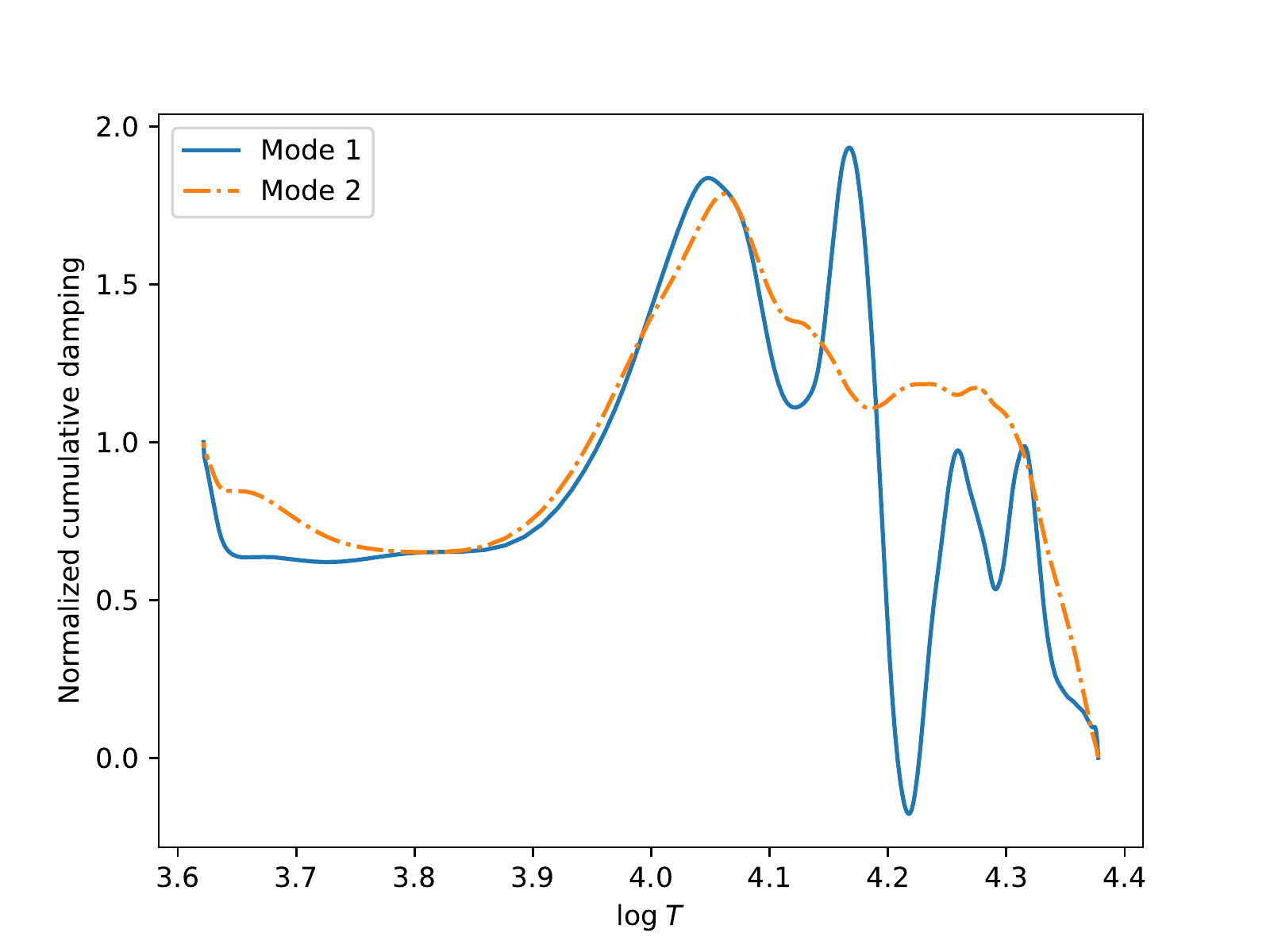}
		\caption{ Cumulative damping computed using Eq.~(\ref{work_integral_form2}) (normalized to the total mode damping rate) starting from the bottom of the simulation, as a function of the logarithm of temperature. Note that due to boundary condition effects, the upper-most layers ($\log T < 3.64$) must be considered with care (see text for details). \label{cumulated_works}}
	\end{center}
\end{figure}

Figure~\ref{cumulated_works} displays the cumulative damping for modes 1 and 2. We stress that when cumulative damping increases (decreases) outward there is a stabilizing effect (destabilizing effect). This convention will be used in the following. For both modes one can distinguish three regions, namely; 
\begin{itemize}
\item The inner quasi-adiabatic region, for which $\log T_0 \gtrsim 4.1$, stabilizes the modes. In this region, however, the cumulative damping is noisy since it exhibits some oscillations. This is due to the Fourier transform (at the modal frequencies) of the pseudo-Lagrangian density which is affected by the non-coherent turbulent fluctuations in this region. This effect is small but as we are considering phase differences it has an important impact on the results.   
\item The atmosphere stabilizes the mode 2 while the cumulative damping is almost constant for mode 1. We also note that near the upper boundary ($\log T_0 \simeq 3.65$) the cumulative damping suddenly increases. This is related to the rapid decrease of the eigenfunctions (see Fig. \ref{identification}, top panel). It is likely an effect of the boundary condition of the simulation.  
\item Finally, the superadiabatic region, which corresponds to the region of hydrogen ionisation ($3.9 \lesssim \log T_0 \lesssim 4.0$), destabilizes. Indeed, this region corresponds to the region where the opacity mechanism related to the ionisation of hydrogen is effective. In the eigenfunctions, this can be seen to be responsible for the rapid variations of the eigen-velocities (see Fig.~\ref{identification}, top panel)
\end{itemize}
Despite of the effect of the non-coherent turbulent fluctuations in the quasi-adiabatic region, the behavioural patterns of the cumulative damping (as described above) are in qualitative agreement with 1D non-adiabatic calculation \citep[e.g.][]{B92a,Belkacem2011}. 

\section{Contributions to the mode damping}
\label{contributions_work_integral}

As shown in the previous section, the measured and computed damping rates are found consistent even if the inherent limitations of the simulation prevents us from getting a perfect match. This encourages us to go further by disentangling the different physical contributions to this damping rate. 

\subsection{The role of gas and turbulent pressure fluctuations}

Let us start by splitting the perturbation of total pressure appearing in \eq{work_integral_form2} as
\begin{align}
\label{split_pressure}
\delta P = \delta P_{\rm g} + \delta P_{\rm t} \, ,
\end{align}
where $\delta P_{\rm g}$ is the perturbation of gas pressure and $\delta P_{\rm t}$ the perturbation of turbulent pressure. Therefore, 
\eq{work_integral_form2} can be rewritten such that 
\begin{align}
\label{cumulated_split_pressure}
\eta  = \eta_{\rm gas} + \eta_{\rm turb} \, , 
\end{align}
where
\begin{align}
\label{contribution_gaz}
\eta_{\rm gas} &=  \frac{\omega_R}{2 E} \int {\rm d}\tau_0 \, \mathcal{I}m\left\{ \frac{\widehat{\delta P_g}}{\rho_0} \frac{\widehat{\delta \rho}^\ast}{\rho_0} \right\} \, , \\
\label{contribution_turb}
\eta_{\rm turb} &=
\frac{\omega_R}{2 E} \int {\rm d}\tau_0 \, \mathcal{I}m\left\{ \frac{\widehat{\delta P_t}}{\rho_0} \frac{\widehat{\delta \rho}^\ast}{\rho_0} \right\} \, .
\end{align}

\begin{figure}[t]
 \begin{center}
   \includegraphics[width=9.8cm]{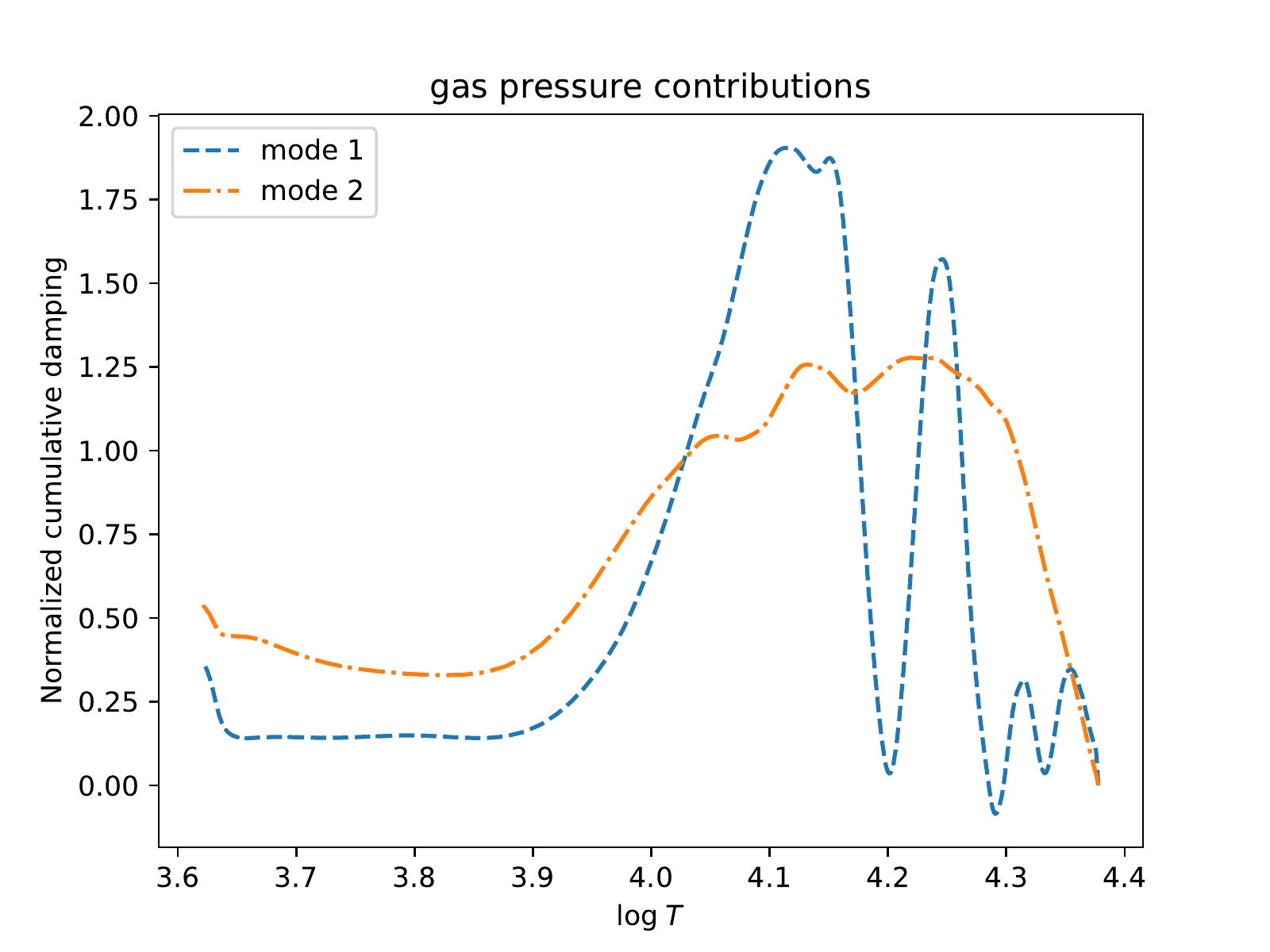}
   \includegraphics[width=9.8cm]{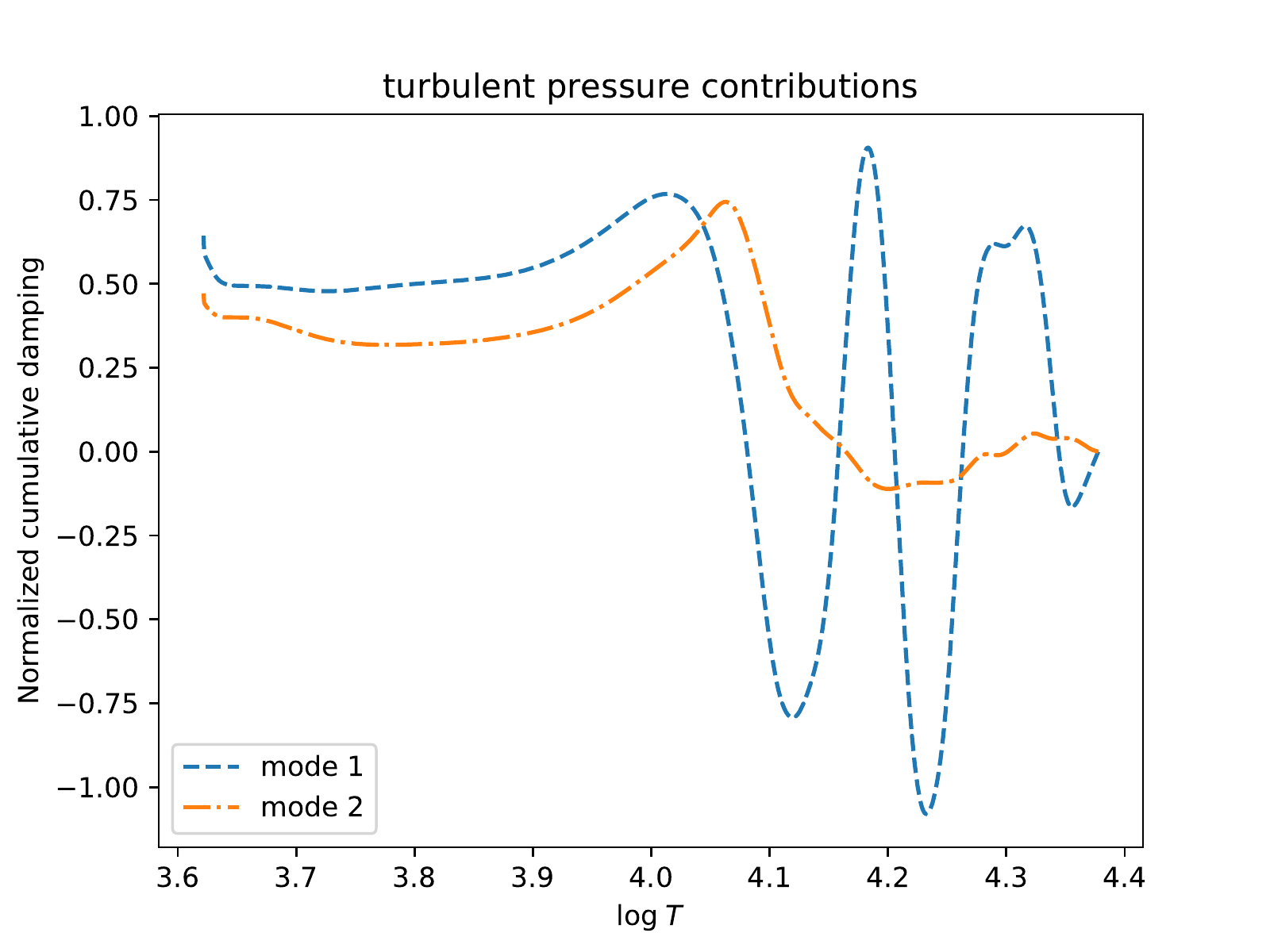}
   \caption{Cumulative work integrals contributions (see \eq{cumulated_split_pressure}) associated with the gas pressure (top panel) and the turbulent pressure (bottom panel), integrated from the bottom of the simulation, as a function of the logarithm of temperature. The total mode damping of each mode is used to normalize the work integrals. Note that due to boundary condition effects, the upper-most layers ($\log T < 3.64$) must be considered with care (see text for details). 
   \label{cumulated_works_contribs_pressure_gaz_turb}}
 \end{center}
\end{figure}

 For these individual contributions to $\eta$, shown in  Fig.~\ref{cumulated_works_contribs_pressure_gaz_turb}, we stress again that for $\log T \gtrsim 4.1$ the results must be considered with care because of the fluctuations of the density perturbation but also of the perturbation of the turbulent pressure.  The Fourier spectra of turbulent pressure are rather noisy, even at modal frequencies, making it difficult to conclude that the signal is dominated by the coherent oscillating signal. Notwithstanding these words of caution, we suggest to distinguish between mainly two regions (for $\log T \lesssim 4.1$). 

First, in the atmospheric layers both modes and both contributions behave the same way, \emph{i.e.} the cumulative damping is almost neutral or only slightly damping. This is due to the fact that, for the contribution associated with the gas pressure, the modal period is much longer than the local thermal time-scale. Consequently, the medium adapts almost instantaneously to any perturbation so that the total energy flux is frozen \citep[see][for a detailed explanation]{Samadi2015}. For the contribution associated with the turbulent pressure, its contribution is very small in the uppermost layers and turns  destabilizing in the inner-layers. This behavior is in qualitative agreement with previous findings by \cite{B92a} and \cite{Sonoi2017}. The latter had shown that the turbulent pressure contribution is mainly controlled by both the Mach number and the ratio of the local convective time-scale to the modal frequency. In the atmospheric layers, both factors are small but increase towards the inner layers to become non-negligible for $\log T \gtrsim  3.9$. 

Secondly, in the super-adiabatic layers (and more precisely for $ 3.9 \lesssim \log T \lesssim 4.0$), the modal period is of the same order of magnitude  as the local thermal time-scale. Consequently, it corresponds to the region where the destabilization effect due to the $\kappa$ mechanism associated  with the ionisation of hydrogen is at work. It also corresponds to the location of the rapid variation of the eigen-velocity shown by Fig.~\ref{identification}. This explains the destabilizing effect of the gas pressure contribution. For the work associated with the turbulent pressure modulation, the situation is equivalent in this region, \emph{i.e.} it destabilizes the modes.  
In contrast, in the inner layers (\emph{i.e.} $\log T \gtrsim 4.0$ for mode 1 and $\log T \gtrsim 4.05$ for mode 2),  excitation by $\delta P_{\rm g}$ competes against damping by $\delta P_{\rm t}$, resulting in a net stabilizing effect for both modes. We note that those behavioural patterns support previous findings using a time-dependent, mixing-length formulation of convection \citep[e.g. Fig. 14 in][]{B92a}. 

\begin{figure}[t]
	\begin{center}
		\includegraphics[width=9.8cm]{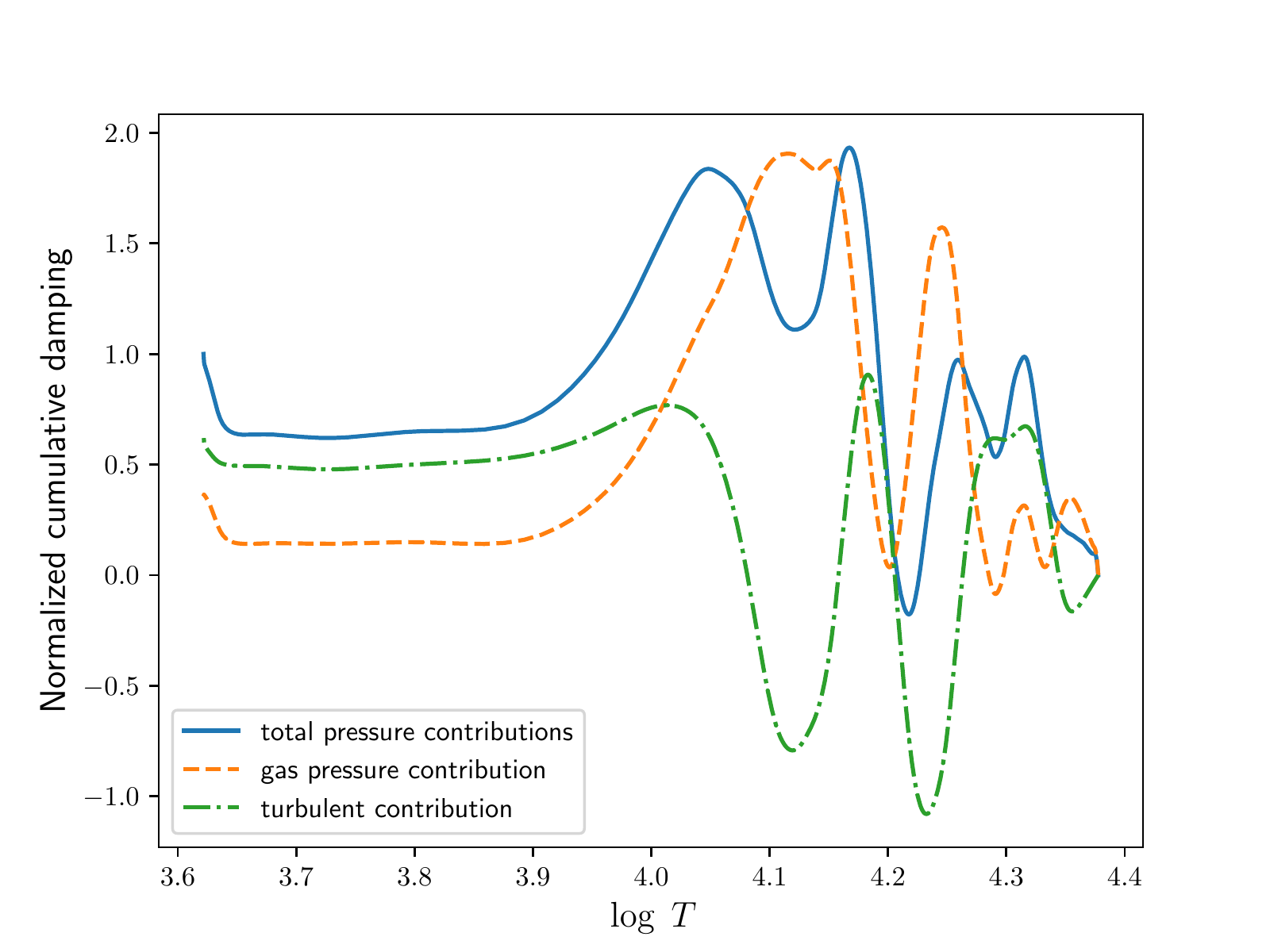}
		\includegraphics[width=9.8cm]{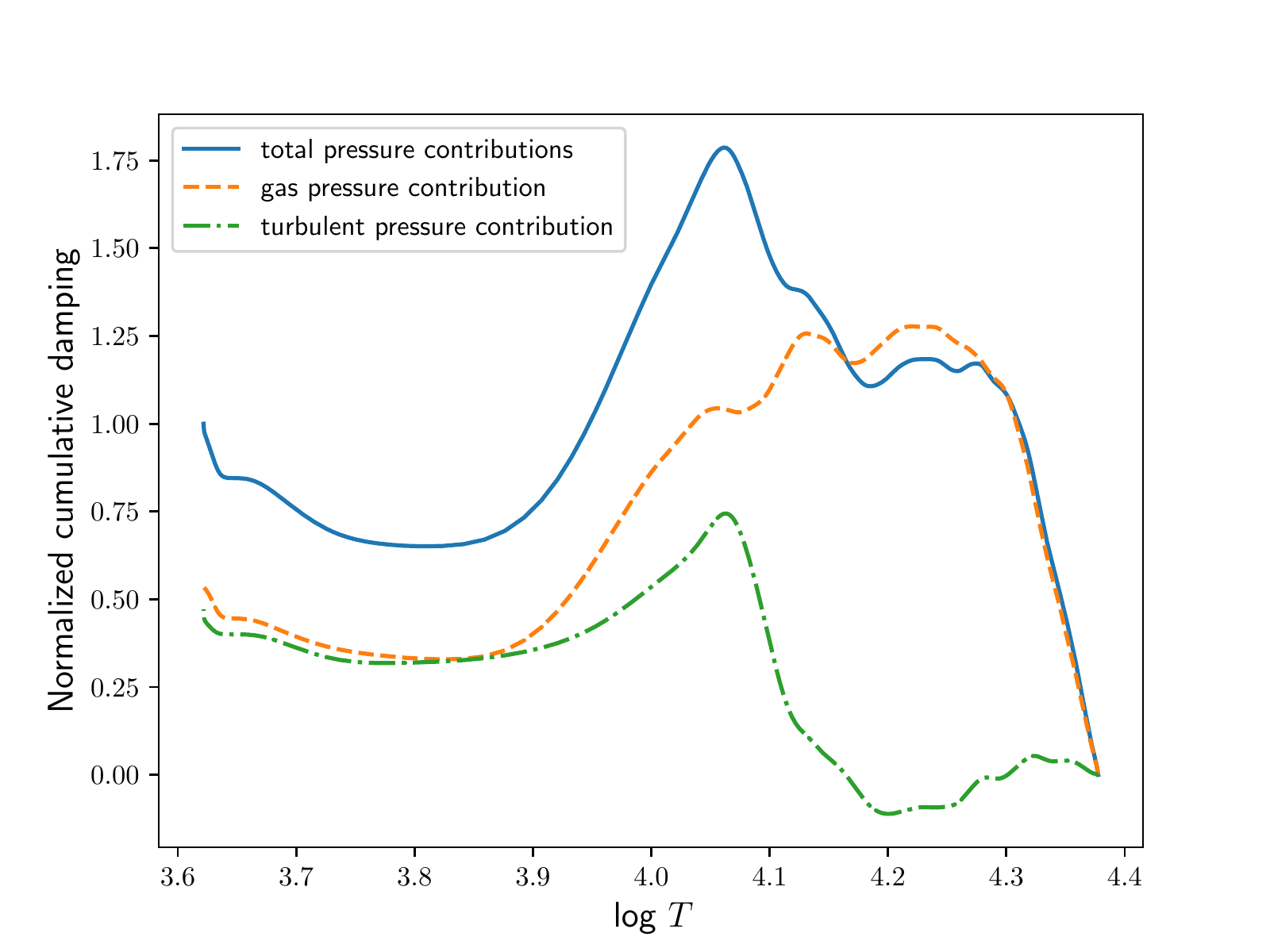}
		\caption{Cumulative work integrals contributions (see \eq{cumulated_split_pressure}) associated with mode 1 (top panel) and mode 2 (bottom panel), starting from the bottom of the simulation, as a function of the logarithm of temperature. The total mode damping of each mode is used to normalize the work integrals. Note that due to boundary condition effects, the upper-most layers ($\log T < 3.64$) must be considered with care (see text for details). \label{cumulated_works_contribs_pressure}}
	\end{center}
\end{figure}

Figures \ref{cumulated_works_contribs_pressure} show the  contributions to the mode damping from the perturbation of the gas pressure (Eq.~\ref{contribution_gaz}) and the turbulent pressure (Eq.~\ref{contribution_turb}). It is worth mentioning that, for the two considered normal modes, both the gas and turbulent pressure contributions have an overall stabilizing effect. For mode 1, the contribution of the turbulent pressure  is slightly  dominant while the two contributions are of the same order of magnitude for mode 2. This is in line with previous findings \citep[see][for a detailed discussion]{Houdek2015} that the role of the perturbation of turbulent pressure is an essential ingredient for stabilizing the modes. However, our result further suggests that the contribution related to the perturbation of the gas pressure also stabilizes the modes and that its contribution becomes dominant, compared to the contribution of turbulent pressure, for high-frequency modes.

\subsection{The contributions to the gas pressure}

To disentangle the various terms that contribute to the perturbation of gas pressure, it is first necessary to use the perturbed equation of state 
\begin{align}
\label{eq_state_perturb}
\frac{\widehat{\delta P_g}}{P_{g,0}} = P_T \, \frac{\widehat{\delta s}}{c_v} + \Gamma_1 \, \frac{\widehat{\delta \rho}}{\rho_0} \, , 
\end{align}
with
\begin{align}
P_T = \left(\Gamma_3-1\right)\,\frac{c_v \rho_0 T_0}{P_{g,0}} \quad \textrm{and} \quad \left(\Gamma_3-1\right) = \left(\derivp{\ln T_0}{\ln \rho_0}\right)_s .
\end{align}
Since the adiabatic part of the gas pressure perturbation (second term of Eq.~\ref{eq_state_perturb}) cancels in the integrand of \eq{work_integral_form2}, the gas component of the damping rate becomes
\begin{align}
\label{work_integral_form3}
\eta_{\rm gas} = \frac{\omega_R}{2 E} \int {\rm d} \tau_0  \,\left(\Gamma_3-1\right) \mathcal{I}m \left(T_0 \widehat{\delta s} \, \frac{\widehat{\delta \rho}^\ast}{\rho_0} \right) \, .
\end{align}
As a consistency check, we computed \eq{work_integral_form3} and recover the values obtained for the gas contribution of the damping as provided by \eq{work_integral_form2}. 

To go further, it is necessary to write down the equation governing the perturbation of entropy. This is obtained from the equation of conservation of the internal energy, which after some manipulations permits us to get to the lowest order (see Appendix.~\ref{energy_conservation_appendix} for details)
\begin{align}
\label{perturb_entropy_final}
i \omega \rho_0 T_0 \, \widehat{\delta s} = 
- \derivp{}{z} \left(\widehat{\delta F}_{{\rm rad},z} + \widehat{\delta F}_{{\rm conv},z} \right) + \widehat{\delta \epsilon} \, , 
\end{align}
where $s$ is the specific entropy, $F_{{\rm rad,z}}$ is the vertical component of the radiative flux,  $F_{{\rm conv,z}}$ is the vertical component of the convective flux defined by 
\begin{align}
\label{def_conv_flux}
\overline{F}_{{\rm conv},z} = \overline{ \rho e^{\prime \prime} u_z^{\prime \prime} + P_g u_z^{\prime \prime}} \, , 
\end{align}
where $e$ is the specific internal energy and the perturbation of dissipation rate of turbulent kinetic energy into heat ($\epsilon$) is defined by 
\begin{align}
\epsilon \equiv \overline{u_j^{\prime\prime} \derivp{ P_g}{x_j} + Q_{\rm diss}} \, , 
\end{align}
with $Q_{\rm diss}$ standing for viscous dissipation. Equation (\ref{perturb_entropy_final}) is the perturbation of the energy equation and is similar to what is obtained from the linear perturbation theory \cite[e.g.][]{Ledoux1958,Ledoux1958b,MAD05}. It emphasizes that the perturbation of gas pressure is the combination of three main physical ingredients, namely the perturbation of the radiative flux, the perturbation of the convective flux and the perturbation of dissipation rate of turbulent kinetic energy into heat.  

Using \eq{work_integral_form3} together with \eq{perturb_entropy_final}, finally gives
\begin{align}
\eta = \eta_{\rm rad} + \eta_{\rm conv} + \eta_{ \epsilon} + \eta_{\rm turb} \, , 
\end{align}
where $\eta_{\rm turb}$ is given by \eq{contribution_turb} and 
\begin{align}
\label{contrib_frad}
\eta_{\rm rad} &= \frac{1}{2 E} \int {\rm d} \tau_0  \,  \left(\Gamma_3-1\right) 
\mathcal{R}e \left(\frac{1}{\rho_0} \derivp{\widehat{\delta F}_{\rm rad}}{z} \frac{\widehat{\delta \rho}^\ast}{\rho_0}\right) \, ,  \\
\label{contrib_fconv}
\eta_{\rm conv} &= \frac{1}{2 E} \int {\rm d} \tau_0  \,  \left(\Gamma_3-1\right) 
\mathcal{R}e \left(\frac{1}{\rho_0} \derivp{\widehat{\delta F}_{\rm conv}}{z} \frac{\widehat{\delta \rho}^\ast}{\rho_0}\right) \, ,  \\
\label{contrib_epsilon}
\eta_{ \epsilon} &= - \frac{1}{2 E} \int {\rm d} \tau_0  \,  \left(\Gamma_3-1\right) 
\mathcal{R}e \left( \frac{\widehat{\delta \epsilon_2}}{\rho_0} \frac{\widehat{\delta \rho}^\ast}{\rho_0}\right) \, .
\end{align}
 The last term ($\eta_{ \epsilon}$) balances the contribution of the turbulent pressure in the quasi-adiabatic region as shown by \cite{Ledoux1958} and \cite{MAD05}. It is thus a non-negligible contribution to the total damping rate but is difficult to estimate directly from the simulation. 
There are two main reasons, one which is method dependent while the other one is related to the physics of dissipation of turbulent kinetic energy. To compute the former in the case of ANTARES would require to evaluate the dissipation through the (Smagorinsky-Lilly type) subgrid scale model \citep[see][and in particular Sect. 2.6 of Mundprecht et al. 2013]{Muthsam2010}. \nocite{Mundprecht2013}  
But this were only a lower limit, since shocks and steep gradients in general would be dealt with by a local, non-linear viscosity which is built into the weighted essentially non-oscillatory scheme used by ANTARES \citep{Muthsam2010}. One of the advantages of the latter is that it aims at minimizing the amount of viscosity added by the scheme to the numerical solution. On the other hand, it is difficult to accurately quantify the exact amount of numerical viscosity introduced this way, at least from a single simulation. A more fundamental, physical problem is that the dissipation of turbulent kinetic energy is dominated by variations of the numerical solution close to the grid scale. Computed values are thus very sensitive to both the numerical method used and to the resolution of the simulation. Since kinetic energy is dissipated down to scales orders of magnitudes smaller than can reasonably be resolved in a simulation of solar convection \citep[a classical result, see also][for references and estimates]{Kupka2017}, any direct computation of this quantity is necessarily inaccurate. To get a better insight would require an expensive series of simulations down to very high resolutions to see if some simple scaling estimates were adequate.

An alternative which provides a first approximation and is available from the present calculations is to estimate $\eta_{\epsilon}$ from its definition, i.e.,
\begin{align}
\label{trick_epsilon}
\eta_{\epsilon} = \eta_{\rm gas} - \eta_{\rm rad} - \eta_{\rm conv} \, .
\end{align}
We use this relation in the following, although we will focus on the radiative and convective flux contributions to the damping rates, \emph{i.e.} \eq{contrib_frad} and \eq{contrib_fconv}, respectively. 

\begin{figure}[t]
  \begin{center}
    \includegraphics[width=9.8cm]{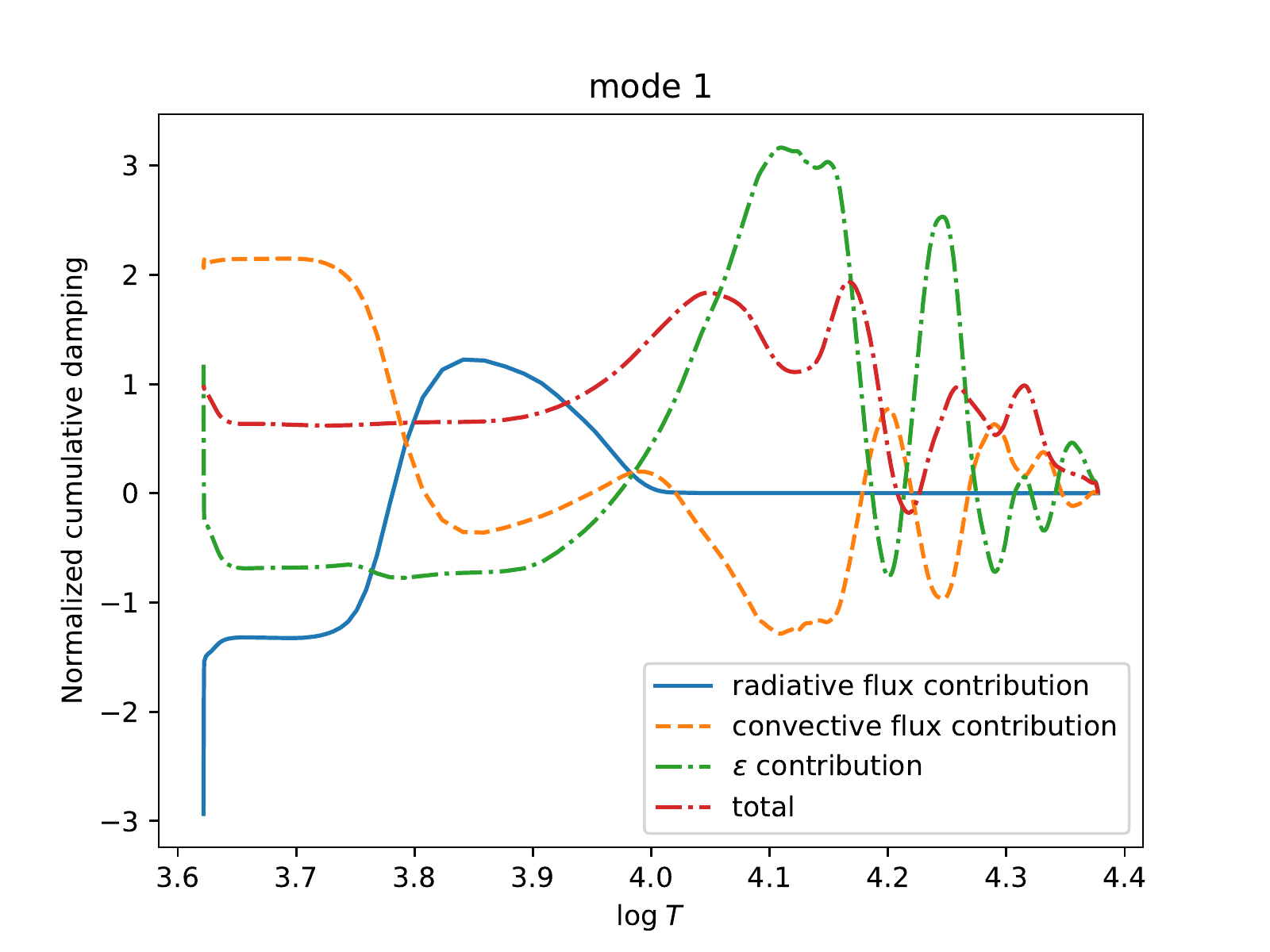}
    \includegraphics[width=9.8cm]{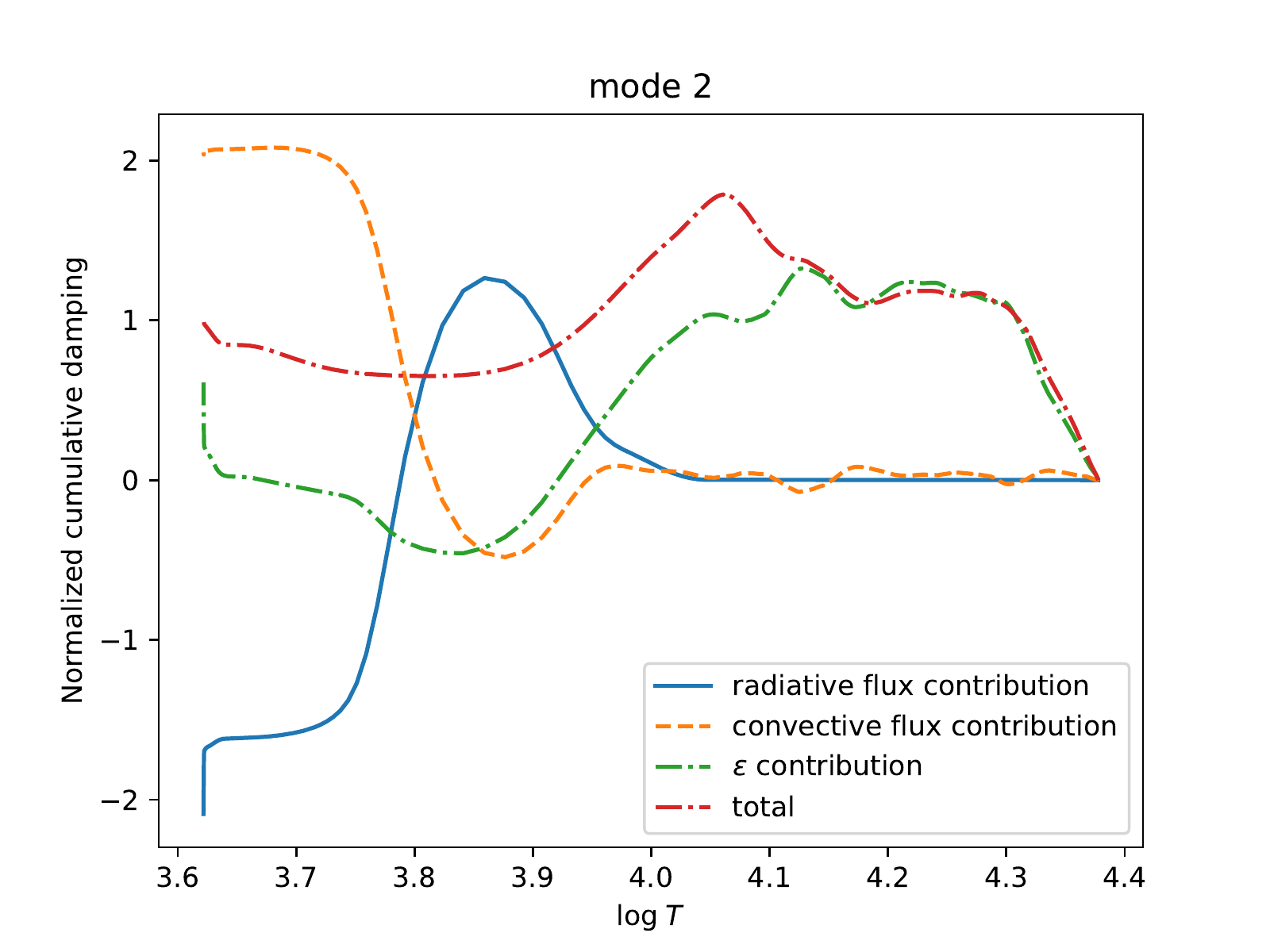}
    \caption{Cumulative work integrals contributions of the gas pressure associated with mode 1 (top panel) and mode 2 (bottom panel), starting from the bottom of the simulation, as a function of the logarithm of temperature. The total mode damping of each mode is used to normalize the work integrals. Note that due to boundary condition effects, the upper-most layers ($\log T < 3.64$) must be considered with care (see text for details). \label{cumulated_works_contribs_pressure2}}
  \end{center}
\end{figure}

Figures \ref{cumulated_works_contribs_pressure2} display the contributions of the divergence of the radiative and convective fluxes to the damping for modes 1 and 2.  For both modes, the behavioral patterns are the same for both the perturbation of  radiative and convective fluxes. In the superadiabatic region, the perturbation of the divergence of the radiative flux stabilizes the modes while the perturbation of the divergence of the convective flux destabilizes them. In the atmospheric layers, the situation is reversed, except for the upper layers, where contributions from both fluxes are close to zero, and the layers at the very top, which are influenced by the boundary conditions. It is worth noting that both contributions are quite large in absolute values but they compensate each other so that the resulting contribution remains small. In addition, if we do not consider the uppermost layers in which the effect of the boundary conditions of the numerical simulation are important, the convective contribution dominates over the radiative one. The total contribution to the damping rate is thus a residual that results from a balance between the two flux divergences 
\cite[see][for a detailed discussion on this issue]{Houdek2015}. Finally, we note that since the contribution of the two fluxes nearly cancels each other, the role of the perturbation of the dissipation rate of turbulent kinetic energy into heat becomes important. This is shown in Figs.~\ref{cumulated_works_contribs_pressure2}. Indeed, these contributions seem to dominate the effect of the perturbation of gas pressure on the mode damping as they essentially destabilize the modes. But as already mentioned, it is difficult to estimate it directly from the numerical simulation, therefore \eq{trick_epsilon} is used to circumvent the problem and as such must be considered with care. Nevertheless, the fact that the contributions of the perturbation of the radiative and convective fluxes nearly cancel each other still emphasizes the essential role of the perturbation of the dissipation rate into heat. 

\section{Concluding remarks}
\label{conclusion}

Using a 3D hydrodynamical simulation of the Sun computed with the ANTARES code and with a time-duration of about 11 hrs, we have shown that it is possible to identify at least three radial normal modes in the simulation, two of which were usable for our purposes. Those modes have been shown to have properties similar to the normal modes of 1D solar models  that happen to have a node  at the bottom of the simulation domain. In contrast,  the amplitudes of simulation modes are found to be much higher than in the Sun due to the relatively very small horizontal area of the simulation. Assuming that the physical background experienced by those modes is realistic enough, we have demonstrated that it is possible to gain some insight into the physics governing the mode damping rates. 

For the two first normal modes of the simulation, it has been possible to investigate the work related to their damping. Except for the quasi-adiabatic region of the simulation, for which the ratio of the mode amplitude compared to the turbulent noise is not large enough, we have been able to exhibit the different regions in which the modes are destabilized and stabilized. Going further, we disentangled the respective role of both the perturbation of the gas and turbulent pressure. From a qualitative point of view our results are in good agreement with previous findings \citep[e.g.][]{B92a,Belkacem2012}. However, from a quantitative point of view, 
it appears that both contributions have an overall stabilizing effect. Indeed, in contrast  to previous results based on time-dependent extensions of the 1D, mixing-length  formulation, the relative contribution associated with the perturbation of the turbulent pressure is not found to be always dominant over the contribution associated with the perturbation of the gas pressure. In addition, it has also been possible to gain insight into the contributions of the perturbation of the divergence of both the radiative and convective fluxes. It appears that those two contributions nearly cancel each other both in the atmospheric and in the  super-adiabatic layers. Indeed, while each of them has an important absolute contribution, the sum of  the two remains small since they tend to cancel each other. However, it appears that while the radiative contribution destabilizes the modes, the convective one stabilizes the mode. The latter is found to be dominant, even if only slightly.

Consequently, we have shown that investigating the properties of the normal modes of a  3D  simulation is of great help to gain some physical insight into the physics of mode damping. From this work, such an approach is found to be feasible. However, the main limitation has been found to be the duration of the simulation. Indeed, it is a crucial point which affects our ability to provide reliable quantitative estimates. A much longer simulation is thus highly desirable. First, it would ensure that the mode of the lowest frequency is resolved, \emph{i.e.} that the linewidth of the mode is higher than one bin in the Fourier domain. It would also ensure that high frequency modes with a very large linewidth could be fitted properly. Second, a longer duration of the simulation is important to improve the ratio between the amplitudes of the normal modes and the turbulent noise in the Fourier spectrum. This is particularly important in the inner-most layers of the simulation. Indeed, mode 2 can already be successfully analysed in detail with the present simulations though of course a longer simulation would be desirable to do the same for mode 1 at $\log T > 4.1$ and possibly also for mode 3. 

We conclude that using 3D hydrodynamical simulations to investigate the physics of mode damping reveals to be a promising approach.  
A drawback is that such an approach is highly demanding in terms of computational efforts because, to get precise and accurate constraints, one would need simulations with a time-duration of several days. This is certainly an objective to attain for our future works on the issue of mode damping. Another important issue that will certainly deserve  further work is related to the top boundary conditions. Indeed, in the very top layers of the simulation the normal modes must be considered with care. This emphasises the need of investigating systematically the influence of boundary conditions on the normal modes of the simulation. 

 Finally, we emphasize that the estimate of the influence of the dissipation on the oscillation remains to be consolidated. Indeed, due to the ENO scheme employed by ANTARES, it is not possible to directly and properly estimate the dissipation within the simulation. A future work dedicated to this issue is thus highly desirable. A possible approach would be to include artificial viscosities in ANTARES and compute the advective fluxes from standard centred stencil. Such a procedure, while being less accurate for the modeling of the dissipation, presents the advantage of permitting to quantify directly its impact on the oscillations and thus to verify that our indirect estimate is valid.

\begin{acknowledgements}
F. Kupka is grateful for support through Austrian Science Fund (FWF) projects P21742-N16, P25229-N27, and P29172-N27. K.B. and R.S. acknowledge support from the ``Programme National de Physique Stellaire'' (PNPS) of CNRS/INSU, France. 
\end{acknowledgements}

\bibliographystyle{aa}

\newpage


\appendix 
\section{Averaged equations}

\subsection{Reynolds and mass (or Favre) averages}
\label{defs_averages}

 Two types of horizontal averages are defined, the first of which
is commonly known as the Reynolds average. In the following we will  approximate the Reynolds average  by the straight horizontal average in the 3D simulation. Therefore, any
 quantity $X$,  is decomposed such that
\begin{align}
X = \overline{X}  + X^\prime \, , 
\end{align}
so that we obviously have 
\begin{align}
\overline{X^\prime} = 0 \, .
\end{align}
The second average is called the mass or Favre average, so that for a quantity $X$, it is defined as
\begin{align}
\widetilde{X} = \frac{\overline{\rho X}}{\overline{ \rho }} \, , 
\end{align}
with 
\begin{align}
X = \widetilde{X} + X^{\prime\prime} \,  .
\end{align}
It immediately follows
\begin{align}
\overline{\rho X^{\prime\prime} } &= 0 \, , \\
\overline{ X^{\prime\prime} } &\neq 0 \, .
\end{align}
From the above equations, it is quite straightforward to derive the following relations
\begin{align}
\label{useful_relations}
\overline{\zeta^{\prime\prime}} &= \overline{ \zeta } - \widetilde{\zeta} = - \frac{\overline{\rho^\prime \zeta }}{\overline{ \rho }} = - \frac{\overline{ \rho^\prime \zeta^{\prime\prime} }}{\overline{ \rho }}\\
\overline{ \rho^\prime \zeta^{\prime\prime} } &= \overline{ \rho^\prime \zeta } = \overline{ \rho^\prime \zeta^{\prime} } \\
\overline{ \rho \zeta \xi } &= 
\overline{ \rho \zeta^{\prime\prime} \xi^{\prime\prime}} 
+ \overline{\rho} \widetilde{\zeta} \,\widetilde{\xi} 
= \overline{\rho} \left[\widetilde{\zeta^{\prime\prime} \xi^{\prime\prime}}+\widetilde{\zeta}\, \widetilde{\xi}\right] \, , 
\end{align}
which will be used in the following. 

\subsection{Mean equations}

The Reynolds averages will be applied to the density and pressure while mass average will be applied to the velocity, temperature and entropy fields. 

\subsubsection{Mass conservation}
\label{mean_mass_conv}

\begin{align}
\derivp{\rho}{t} + \derivp{}{x_j} \left(\rho u_j\right) = 0 \, , 
\end{align}
where $\rho$ stands for the density and $u_j$ the $j$ component of the velocity field. After averaging it gives
\begin{align}
\label{mass_average_general}
\derivp{\overline{\rho}}{t} + \derivp{}{x_j} \left(\overline{\rho} \,\widetilde{u}_j\right) = 0 \, .
\end{align}
To go further, we assume that there is no large scale horizontal flow, like meridional circulation, so that there is no mean horizontal momentum flux. Therefore, 
Eq.~(\ref{mass_average_general}) reduces to 
\begin{align}
\label{mass_average_general2}
\derivp{\overline{\rho}}{t} + \derivp{}{z} \left(\overline{\rho} \,\widetilde{u}_z\right) = 0 \, .
\end{align}
If we now introduce the pseudo-Lagrangian derivative as 
\begin{align}
\deriv{}{t} = \derivp{}{t} + \widetilde{u}_z \derivp{}{z} \, , 
\end{align}
we finally get
\begin{align}
\deriv{\overline{\rho}}{t} + \overline{\rho}\derivp{\widetilde{u}_z}{z} = 0 \, .
\end{align}
This equation is the same as used by Nordlund \& Stein (2001).

\subsubsection{Momentum conservation}

\begin{align}
\label{app_momentum_eq0}
\derivp{\rho u_j}{t} + \derivp{}{x_k} \left(\rho u_j u_k\right) = -\derivp{P_g}{x_j} - \rho g_j \, , 
\end{align}
with $P_g$ the gas pressure and $g$ the gravitational acceleration. After averaging, one gets 
\begin{align}
\derivp{\overline{\rho}\, \widetilde{u}_j}{t} + \derivp{}{x_k} \left(\overline{\rho} \, \widetilde{u}_j \widetilde{u}_k \right) 
+ \derivp{}{x_k} \overline{\rho \, u_j^{\prime \prime} u_k^{\prime \prime} }  = -\derivp{\overline{P}_g}{x_j} - \overline{\rho} g_j \, .
\end{align}
Using the same approximation as for Eq.~(\ref{mass_average_general2}), it reduces to 
\begin{align}
\derivp{\overline{\rho}\, \widetilde{u}_j}{t} + \derivp{}{z} \left(\overline{\rho} \, \widetilde{u}_j \widetilde{u}_z \right) 
+ \derivp{}{z} \overline{\rho \, u_j^{\prime \prime} u_z^{\prime \prime}}  = -\derivp{\overline{P}_g}{x_j} - \overline{\rho} g_j \, .
\end{align}
Because we are interested in radial modes, we consider $j=z$ to obtain 
\begin{align}
\derivp{\overline{\rho}\, \widetilde{u}_z}{t} + \derivp{}{z} \left(\overline{\rho} \, \widetilde{u}_z^2 \right) 
+ \derivp{}{z} \overline{\rho \, u_z^{\prime \prime 2} }  = -\derivp{\overline{P}_g}{z} - \overline{\rho} g \, .
\end{align}
In the pseudo-Lagrangian frame this simplifies to 
\begin{align}
\label{momentum_final}
\overline{\rho} \deriv{\widetilde{u}_z}{t} = -\derivp{\overline{P}_g}{z} -\derivp{\overline{P}_t}{z} - \overline{\rho} g \, , 
\end{align}
with 
\begin{align}
\overline{P}_t = \overline{\rho \, u_z^{\prime \prime 2}} \, .
\end{align}
From Eq.~(\ref{momentum_final}), the stationary solution in the pseudo-Lagragian frame reads
\begin{align}
\label{momentum_stationary}
-\derivp{\left< \overline{P}_g \right>_t}{z} -\derivp{\left< \overline{P}_t \right>_t }{z} - \left< \overline{\rho} \right>_t g = 0 \, , 
\end{align}
where $\left<\right>_t$ denotes the temporal average. 
Thus, one can use Eqs.~\ref{momentum_final} and \ref{momentum_stationary} to finally get
\begin{align}
\label{momentum_wave}
\overline{\rho} \deriv{\widetilde{u}_z}{t} = -\derivp{\delta P_g}{z} -\derivp{\delta P_t}{z} - \delta \rho g \, , 
\end{align}
where $\delta X \equiv \overline{X}-\left< \overline{X}\right>_t$. 

\subsubsection{Energy equation}
\label{energy_conservation_appendix}

To go further, and disentangle the various contributions of the perturbation of gas pressure, it is necessary to write the equation governing the perturbation of entropy. To this end, 
\begin{align}
\label{internal_energy}
\derivp{}{t}\left(\rho e\right) + \derivp{}{x_j}\left(\rho e u_j\right) = - P_g \derivp{u_j}{x_j} - \derivp{F_{{\rm rad},j}}{x_j} + Q_{\rm diss} \, , 
\end{align}
where $e$ is the specific internal energy, $F_{{\rm rad}}$ is the radiative flux, and $Q_{\rm diss}$ stands for viscous dissipation. After averaging, Eq.~(\ref{internal_energy}) becomes 
\begin{align}
\label{internal_energy_mean_tmp}
&\derivp{}{t}\left(\overline{\rho} \, \widetilde{e}\right) 
+ \derivp{}{x_j}\left[\rho \, \overline{e^{\prime\prime} u_j^{\prime\prime}} 
+ \overline{\rho} \, \widetilde{e}\, \widetilde{u}_j\right] \nonumber \\
&= - \overline{P_g} \, \derivp{\widetilde{u}_j}{x_j} -  \derivp{\overline{P_g u_j^{\prime\prime}}}{x_j} 
+ \overline{u_j^{\prime\prime} \derivp{ P_g}{x_j}}
- \derivp{\overline{F_{{\rm rad},j}}}{x_j} + \overline{Q_{\rm diss}} \, .
\end{align}
Finally, after some rearrangements the pseudo-Lagrangian
energy equation is 
\begin{align}
\label{internal_energy_mean}
\overline{\rho} \deriv{\widetilde{e}}{t} = 
- \derivp{}{z} \left(\overline{F}_{{\rm rad},z} + \overline{F}_{{\rm conv},z}\right) 
- \overline{P_g} \, \derivp{\widetilde{u}_j}{x_j} + \overline{u_j^{\prime\prime} \derivp{ P_g}{x_j}} + \overline{Q_{\rm diss}} \, ,
\end{align}
with 
\begin{align}
\label{def_conv_flux}
\overline{F}_{{\rm conv},z} = \overline{\rho e^{\prime \prime} u_z^{\prime \prime} + P_g u_z^{\prime \prime}} \, .
\end{align}
Now, to the lowest order, we use the relation
\begin{align}
\label{entropy_low_order}
 \deriv{\widetilde{e}}{t} = \overline{T} \, \deriv{\widetilde{s}}{t} 
- \frac{\overline{P}_g}{\overline{\rho}^2} \, \deriv{\overline{\rho}}{t} 
= \overline{T} \, \deriv{\widetilde{s}}{t} - \frac{\overline{P}_g}{\overline{\rho}} \derivp{\widetilde{u}_j}{x_j} \, , 
\end{align}
where $T$ is the temperature and $s$ the specific entropy. Consequently, Eq.~(\ref{internal_energy_mean}) becomes to the lowest order
\begin{align}
\label{entropy_final}
\overline{\rho} \overline{T} \, \deriv{\widetilde{s}}{t} = 
- \derivp{}{z} \left(\overline{F}_{{\rm rad},z} + \overline{F}_{{\rm conv},z}\right)  
+ \overline{\epsilon} \, , 
\end{align}
where we have defined
\begin{align}
\label{def_epsilon_app}
\overline{\epsilon} \equiv \overline{ u_j^{\prime\prime} \derivp{ P_g}{x_j} + Q_{\rm diss}} \, .
\end{align}
Finally, we note that while contributions by viscosity can be neglected in \eq{app_momentum_eq0} and thus also in \eq{momentum_wave}, the same does not hold
true for \eq{internal_energy} and hence \eq{entropy_final}-\eq{def_epsilon_app}, as has been critically discussed by \cite{Canuto1997b}. Hence, in \cite{Canuto1997} the viscous flux was dropped in the dynamical equation for the large scale velocity field, whereas the dissipation rate of turbulent kinetic energy into heat, $\epsilon$, was kept in the dynamical equations for the Reynolds stresses and for the temperature field, i.e., for the energy equation.

\end{document}